# 3D front tip fields in creeping solids under constraint effects: a higher-order asymptotic solution

Weichen Kong [a], Yanwei Dai [b,*], Yinghua Liu [a,*]

[a] Department of Engineering Mechanics, AML, Tsinghua University, Beijing 100084, China

[b] Institute of Electronics Packaging Technology and Reliability, Faculty of Materials and Manufacturing, Beijing University of Technology, Beijing 100124, China

[*] Corresponding authors:

E-mail addresses: ywdai@bjut.edu.cn (Y. Dai), yhliu@mail.tsinghua.edu.cn (Y. Liu).

# Abstract

As one of the most important topics studied in creep fracture mechanics, mechanics fields at three-dimensional (3D) sharp V-notches and crack tip have drawn tremendous attentions. With many years efforts on constraint theory developed in creeping solids, there still seems dense fog on how in-plane and out-of-plane constraint effects are interacted for 3D sharp V-notch and crack in creeping solids. To shed lights on this topic, a 3D higher-order termed solution for sharp V-notches in creeping materials subjected to mode I loading is established by introducing the out-of-plane factor $T_z$, which is the out-of-plane stress divided by the sum of in-plane normal stress. The solution can naturally be degenerated to a 3D crack. Based on the 3D higher-order term solution, a new fracture parameter $A_2^T$ is proposed and combined with $T_z$ to characterize 3D constraint effect. It is found that the stress exponents and angular distribution of higher-order term for 3D notches and cracks are highly related to $T_z$. The proposed higher order termed solutions show better agreement with the FEA results than the 3D leading-term and 2D two-term solutions, especially for smaller notch angles and ligament width. Moreover, the presented 3D constraint theory shows that effects of $T_z$ and $A_2^T$ are highly interlinked rather than simply separated. It implies that the 3D constraint level may be significantly influenced by $T_z$. The 3D mathematical solutions discussed in this paper could enhance the understanding of the 3D effect and has the potential to explain the 3D constraint effect on the notches and cracks under creep conditions.

# 1. Introduction

Three-dimensional (3D) fracture-related issues are possibly the most challenging and charming topic in fracture mechanics. Understanding, accurate and reasonable descriptions of the 3D crack or notch tip fields are preliminary for further engineering applications of those built-up fracture mechanics frameworks. Due to higher requirements on global trends of carbon neutrality, energy saving, and environment protection, large amounts of mechanical components, electronic devices, and civil nuclear infrastructures will be operated under higher temperatures to promote energy conversion and utilization. For those components serviced at elevated temperatures under harsh and extreme conditions, viscoplastic damage or creep damage and cracking need to be evaluated efficiently and accurately in structural integrity assessment (Ainsworth, 2006; James et al., 2020), especially for those 3D crack and notch containing components.

On 3D crack fronts, researchers have been working on this for a long time, and it is still in process. Earlier studies always focused on the calculation and distribution of stress intensity factor (SIF) and energy release rate of 3D cracks through numerical computations, e.g., boundary-integral method (Weaver, 1977), body-force method (Lee and Keer, 1986), and domain integral (Shih et al., 1986). Rigorous theoretical analysis was analyzed through weight function theory (Rice, 1989). Further, the application of advanced observation technology makes it possible to quantify the displacement fields of 3D cracks (Mostafavi et al., 2013; Tonge et al., 2020). The 3D fracture of new biomaterials has also attracted increasing attention (Meng et al., 2022). It is widely accepted that the characteristics of 3D cracks are different from those in 2D conditions. For example, there is a pronounced loss of HRR dominance along the crack front for 3D elastoplastic plates (Nakamura and Parks, 1990). The fracture behavior of 3D cracks is related to the degree of plane strain ($D_{p\varepsilon} = \sigma_{zz}/\left[\nu\left(\sigma_{xx}+\sigma_{yy}\right)\right]$)(Kwon and Sun, 2000), which is closely similar to the out-of-plane effect. Those investigations provide foundations for understanding and evaluating 3D fracture problems.

On 3D crack in elastic-plastic and creeping solids, there are some relevant works studying the asymptotic solutions of tip fields. Guo and his co-authors have done a series of innovative works in the tip fields of 3D crack under out-of-plane constraint effects since the 1990s (Guo, 1993a, b, 1995). The out-of-plane factor $T_z$ was introduced into the governing equation of the crack tip fields in power-law plastic solids and discussed comprehensively to quantify the out-of-plane constraint effect of 3D crack front. Xiang et al. (2011) analyzed the crack tip fields in creeping solids based on $T_z$, and employed $Q^*$ to characterize in-plane constraint effect where detailed analyses were also carried out (Xiang and Guo, 2013). Recently, the second-order term theory was used to establish the in-plane constraint parameter $A_\mathrm{T}$ for 3D cases. Herein, $A_\mathrm{T}$ is combined with $T_z$ to characterize both in-plane and out-of-plane constraint effects in plastic (Cui and Guo, 2019) and creep (Cui and Guo, 2020) solids. Both $Q^*$ and $A_\mathrm{T}$ are obtained based on the first-order term which is under the effect of $T_z$. Moreover, different constraint parameters in describing 3D creep crack front were discussed by Matvienko et al. (2013) through 3D numerical analyses.

From the perspective of fracture mechanics bearing creep and elastoplastic conditions, the HRR field (Hutchinson, 1968; Rice and Rosengren, 1968) and the RR field (Riedel and Rice, 1980) have been built up with facture parameters $J$-integral and $C(t)$-integral. However, the HRR-dominant or RR-dominant areas are limited for 3D cracked and notched components with finite size (Li et al., 2000; Xiang et al., 2011). Higher-order term solutions shall be developed to characterize the 3D crack and notch apexes with finite sizes (Chao and Lam, 2009; Dai et al., 2021). The constraint effect, which reflects the geometry of specimens and the loading types, has a significant influence on the tip fields (Dai et al., 2021; Kong et al., 2022a), failure behaviors (Cui and Guo, 2022; Davies et al., 2009; Ma et al., 2016; Zhao et al., 2023), and evaluation (Dai et al., 2020a; Kong et al., 2022b) of cracks or notches under creep condition. In-plane and out-of-plane stress states were considered nonnegligible for computing constraint effect of the 3D notch and crack. In order to calibrate the constraint effect of the 3D crack or notch apexes, the intrinsic relation between the in-plane stress state and the out-of-plane stress state needs to be presented. Characterizing the constraint effect of 3D notches and cracks is an increasingly important topic for 3D fracture issues.

Based on Riedel and Rice (1980) (famed RR field), a single parameter $C(t)$ is proposed to characterize the tip field. Afterward, the two-parameter solutions considering higher-order terms and the in-plane constraint effect for cracks under creep conditions were developed, such as $C^*-A_2$ and $C^*-Q$ solutions (Budden and Ainsworth, 1999; Chao et al., 2001; Nguyen et al., 2000). Recently, mixed-mode tip fields considering higher-order terms are discussed by Loghin and Joseph (2020) and Dai et al. (2020b). Further, the out-of-plane effect is considered to establish the tip field of 3D cracks under creep conditions by Xiang et al. (2011) and, more recently, by Cui and Guo (2020).

For notches, Kuang and Xu (1987) analyzed the stress field and strain field of sharp V-notch through FEM. Unlike cracks, the singularity behavior and tip fields of sharp V-notches in creeping material depend on notches angels, which has been discussed by Zhu et al. (2011). Lazzarin and Zappalorto (2012) established the 3D stress field of notches based on the generalized plane strain hypothesis. The method was extended to the elastic-plastic cases for 3D notches (Lazzarin et al., 2015). Moreover, the stress and strain fields of sharp V-notches in creeping solids are also discussed by Gallo et al. (2016) and Dai et al. (2019) theoretically and numerically. On 3D sharp V-notch in power-law creep solids, a 3D leading-term asymptotic solution considering the out-of-plane effect was recently reported by Kong and coworkers (Kong et al., 2022a).

In-plane and out-of-plane constraint effects were always considered widely to describe the crack or notch front of 3D crack and notch fronts considering constraint effects. On the one hand, for the in-plane effect, a load-independent parameter $R^*$ was proposed to quantify in-plane constraint under creep conditions (Tan et al., 2014). Dai et al. (2021) adopted $A_2$-term to show the in-plane constraint effect on stress fields of sharp V-notches in creeping solids. Otherwise, Zhao et al. (2023) used the $Q$-type parameters to describe the in-plane constraint effect when studying the creep crack growth behavior. On the other hand, the out-of-plane factor $T_z$ was used to characterize the out-of-plane constraint for 3D sharp V-notches and cracks for creep solids. Kong et al. (2022b) adopted $T_z$ to investigate the out-of-plane effect on the stress fields of 3D sharp V-notches for creep materials. More recently, Cui and Guo (2022) studied the effect of $T_z$ on crack growth in power-law creep solids.

Different from those studies considering the in-plane effect and out-of-plane effect separately, some investigations are also proposed to unify the in-plane and out-of-plane constraint effects. Ma et al. (2016) proposed a unified parameter $A_c$ based on the creep zone size to unify the effect of in-plane and out-of-plane constraints on creep crack growth. Tonge et al. (2020) used synchrotron X-ray tomography with in-situ loading to obtain the plastic zone size to characterize the effect of unified constraint. Cui and Guo (2020) adopted $T_z$ and $A_T$ to characterize out-of-plane and in-plane constraint effects, respectively. The second-order term, i.e., $A_T$-term, is based on the second-order solution under the plane-strain condition.

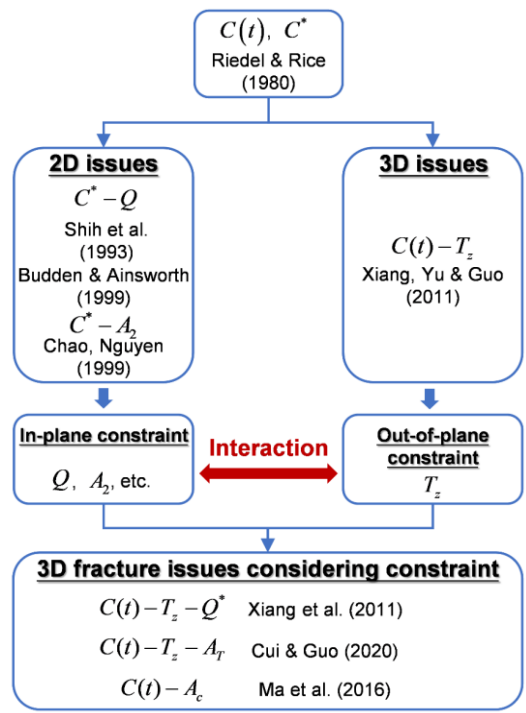

Fig. 1 The characteristic parameters and constraint effects for creep cracks

Dating back to the developing mainstream of the constraint theory for creep crack, as shown in Fig. 1, it is still an unsolved problem to integrate all the constraint effects and introduce it into the crack or notch tip fields. Before that, there raises the question of how and what in-plane and out-of-plane parts will be interacted and interlinked for 3D fracture issues. Since most of those investigations on constraint are generally analyzed based on the plane problem, the in-plane constraint effect is always realized before the out-of-plane constraint, and those studies naturally accept that constraint could be characterized by in-plane part by neglecting out-of-plane part. As the formulation of in-plane and out-of-plane constraints makes the application more complicated, some investigations tried to combine the two constraint effects as a unified one. Relevant researchers put forward the concept of unified constraint such as Ma et al. (2016) and Tonge et al. (2020). However, how the out-of-plane constraint affects the in-plane constraint or how these two constraint effects interlinked are still not answered, which is still a crucial gap in the understanding of constraints in 3D fracture problems.

Hence, in the present paper, the mode I sharp V-notch in power-law creeping solids, which can be degenerated into the 3D crack, will be taken as a general configuration in analysis. Moreover, $T_z$ will be introduced into the higher-order governing equation through hierarchy order asymptotic analysis. Accordingly, a more comprehensive and accurate 3D tip field with rigorously theoretical foundation supported solution will be established. In detail, the rigorous analytical solutions of 3D creep crack or notch tip fields considering higher-order terms are presented. The stress exponents and distribution function of 3D creep cracks or sharp V-notch were systematically studied. Especially, the higher order term solutions of 3D creep crack and sharp V-notch were obtained analytically. With the obtained 3D higher-order term solutions, i.e., the $K_\mathrm{N}^\mathrm{T} - A_2^\mathrm{T}$ solution, 3D constraint effect was presented. The presented 3D constraint theory can quantify all constraint effects reasonably, as well as deepen the understanding of constraint effects in 3D fracture issues. Most importantly, the solution will give a very clear explanation on the relation between in-plane constraint and out-of-plane constraints from the perspective of asymptotic analysis. The correlation between the in-plane and out-of-plane effects was also revealed and discussed systematically through this investigation.

## 2. Formulation

To investigate the eigenvalue of the 3D front tip fields in creeping solids, the following formulation analysis is based on rigorously higher order asymptotic analysis. As the 3D creep crack can be degenerated from the 3D sharp V-notch tip fields only by setting notch angle to be zero, hence the following analysis is naturally valid and applicable for 3D creep crack front conditions. For the convenience of illustrating, the fact that the following solutions are also applicable for the 3D creep crack tip fields will not be emphasized again in the following analysis.

### 2.1 Basic governing equations of 3D sharp V-notch problems

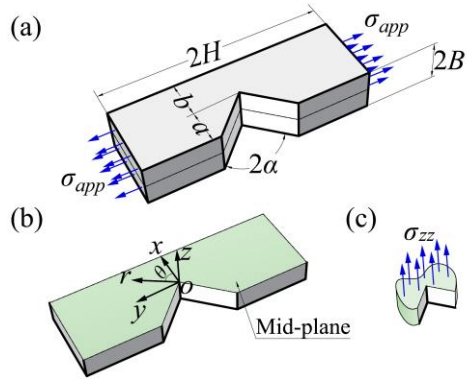

Fig. 2 (a) Specimen geometry and loading (b) Coordinate systems at Mid-plane (c) Out-of-plane stress

A 3D sharp V-notch in creeping materials under mode I loading is considered as shown in Fig. 2a, where the notch angle, thickness, and height of the specimen are denoted as $2\alpha$, $2B$, and $2H$, respectively. The notch depth is denoted as $a$, and the width of the ligament is $b$. For the convenience of analysis, a cartesian coordinate system and a cylindrical coordinate system are established at the notch front. As shown in Fig. 2b, the coordinate origin $O$ is set at the notch tip of the mid-plane of the specimen. The x-axis coincides with the angular bisector of the notch, and the z-axis is perpendicular to the mid-plane.

To obtain the tip fields of 3D sharp V-notches, the 3D governing equations for power-law creeping materials are needed. The constitutive equation is written as

$$\dot{\varepsilon}_{ij} = \frac{1+\nu}{E}\dot{S}_{ij} + \frac{1-2\nu}{3E}\dot{\sigma}_{kk}\delta_{ij} + \frac{3}{2}\dot{\varepsilon}_0\left(\frac{\sigma_e}{\sigma_0}\right)^{n-1}\frac{S_{ij}}{\sigma_0} \quad (1)$$

$$S_{ij} = \sigma_{ij} - \frac{\sigma_{kk}}{3}\delta_{ij} \quad (2)$$

$$\sigma_e^2 = \frac{3}{2}S_{ij}S_{ij} \quad (3)$$

in which $\dot{\varepsilon}_{ij}$, $S_{ij}$, $\dot{\sigma}_{kk}$, $\sigma_e$, $\delta_{ij}$ are strain rate, deviatoric stress, hydrostatic stress rate, Mises equivalent stress, and Kronecker delta, respectively. $E$, $\nu$, $\dot{\varepsilon}_0$, $\sigma_0$, and $n$ represent Young's modulus, Poisson's ratio, reference strain rate, reference stress, and creep exponent, respectively. A dot over the quantities represents the differential of creep time. The subscripts $i, j$ take $r, \theta, z$ for quantities in a cylindrical coordinate system shown in Fig. 2b.

The equilibrium equations in rate form for a 3D problem without body forces are written below in the tensor form

$$\dot{\sigma}_{ij,j} = 0 \quad (4)$$

The following theoretical analysis takes stress function as a fundamental variable. The relation between stress components and stress function is given as

$$\sigma_{ij} = e_{ink}e_{jml}\varphi_{mn,kl} \quad (5)$$

The Maxwell stress function ($\chi_1 = \varphi_{11}, \chi_2 = \varphi_{22}, \chi_3 = \varphi_{33}$) is adopted.

The relation between strain and displacement is written as

$$\dot{\varepsilon}_{ij} = \frac{1}{2}\left(\dot{u}_{i,j} + \dot{u}_{j,i}\right) \quad (6)$$

The strain components should obey the strain compatibility equation:

$$e_{mjk}e_{nil}\dot{\varepsilon}_{ij,kl} = 0 \quad (7)$$

where $e_{ijk}$ is the Eddington tensor, i.e., $e_{ijk} = (i-j)(j-k)(k-i)/2$.

## 2.2 Stress exponent characteristics

Before stress exponent characteristics analysis, here we should know that there are some basic assumptions for the asymptotic solution. One important assumption is the small strain deformation assumption. The second assumption is the creep predominant assumption. It represents that the creep strain is enclosed around the entire notch or crack tip fields, which also indicates that the creep strain is greater than that of elastic strain. Another important assumption is that all the tip field components, such as stress components and strain components, shall be expressed in a series of expansion form.

With those assumptions, the stress components of a 3D sharp V-notch, which considers the out-of-plane effect, are given as the following form without loss of generality:

$$\sigma_{ij}(r,\theta,z,t) = \sum_{\beta} A_{\beta}^{\mathrm{T}}(z,t) r^{s_{ij}^{(\beta)}(T^*(z))} \tilde{\sigma}_{ij}^{(\beta)}(\theta, T_z) \tag{8}$$

in which stress exponent $s_{ij}^{(\beta)}(T^*(z))$ is functions of $T^*(z)$, and stress angular distribution function $\tilde{\sigma}_{ij}^{(\beta)}$ is dependent on both $\theta$ and $T_z$. $A_{\beta}^{\mathrm{T}}(z,t)$ is the amplitude function which is affected by creep time. $\beta$ here represents $\beta$-th order term of stress. The subscript in Eq. (8) does not include a summation convention. Note that the stress exponent and distribution functions of the stress field are related to the out-of-plane factor $T_z$. The introduction and characteristics of the out-of-plane factor $T_z$ are discussed as follows.

It has been widely recognized and validated that the out-of-plane effect plays an essential role in 3D fracture problems theoretically (Cui and Guo, 2022; Kong et al., 2022b; Matvienko et al., 2013; Mostafavi et al., 2011; Qian et al., 2014; Shlyannikov et al., 2011; Sun et al., 2022) and experimentally (Li et al., 2017; O'Connor et al., 2022; Tan et al., 2013; Tonge et al., 2020). In most of the articles mentioned above, the out-of-plane factor $T_z$ is employed in the following form:

$$T_z(t) = \frac{\sigma_{33}}{\sigma_{11} + \sigma_{22}} \tag{9}$$

in which $\sigma_{33}$ is the out-of-plane normal stress $\sigma_{zz}$ shown in Fig. 2c. The out-of-plane factor $T_z$ is successfully used to analyze 3D crack tip fields for elastoplastic materials (Guo, 1993a). Recently, we have reported the out-of-plane effect on the fracture characteristics of 3D sharp V-notch in power-law creeping materials (Kong et al., 2022a). In the present article, the power-law creeping condition is considered. Hence, the out-of-plane factor $T_z(t)$ is time-dependent.

For sharp V-notch under creep conditions, it is necessary to clarify the following hypotheses about the out-of-plane factor $T_z(t)$.

$$\begin{cases} T_z = T_z(r,\theta,z,t) \\ \lim_{r \to 0} T_z = T(z,t) \end{cases} \quad (10)$$

Specifically, the out-of-plane factor is independent of $r$ and $\theta$ for a perfect sharp V-notch in a cylindrical coordinate system when $r$ approaches 0. It is reasonable to consider $T_z(r,\theta,z,t)$ being within $[0, 0.5]$ for creep cases near the notch front for conditions without out-of-plane loads. While for elastic cases, $T_z(r,\theta,z,t)$ distributes in $[0, v]$. The range of $T_z(r,\theta,z,t)$ under creep condition is consistent with the $T_z$ defined by Guo (Guo, 1993a).

Moreover, as $T_z(r,\theta,z,t)$ is a parameter characterizing 3D stress conditions, it is reasonable to consider $T_z$ to be independent of creep time for extensive creep. Specifically, $T_z(r,\theta,z,t)$ approaches $T_z^*(r,\theta,z)$ for long-term creep. As a result, $T(z,t)$ approaches $T^*(z)$ when creep time is sufficiently long. A detailed discussion is reported in our previous work (Kong et al., 2022a).

Similar to the stress components in Eq. (8), the $\beta$-th order term of Maxwell stress function can be written as (without summation):

$$\chi_i^{(\beta)} = D_\beta^T(t) r^{\lambda_i^{(\beta)}(T^*(z))} \tilde{\chi}_i^{(\beta)}(\theta, T_z) \quad (11)$$

in which $\lambda_i^{(\beta)}(T^*(z))$ and $\tilde{\chi}_i^{(\beta)}(\theta, T_z)$ are stress exponent and angular distribution function related to $T^*(z)$ and $T_z$, respectively. It should be noted that this paper focuses on extensive creep conditions. Hence, we assume the fracture characteristics depend on the long-term creep parameters $T^*(z)$ and $T_z$. After expending the stress components in the cylindrical coordinate, the relationship between the Maxwell stress function exponents can be concluded:

$$\lambda_1^{(\beta)} = \lambda_2^{(\beta)} = \lambda_3^{(\beta)} = \lambda^{(\beta)}\left(T^*(z)\right) \tag{12}$$

The stress exponents of the stress components can also be concluded as follows:

$$\begin{cases} \sigma_{xx}^{(\beta)}, \sigma_{yy}^{(\beta)}, \sigma_{xy}^{(\beta)}, \sigma_{zz}^{(\beta)} \sim r^{\lambda^{(\beta)}(T^*(z))-2} \sim r^{s^{(\beta)}(T^*(z))} \\ \sigma_{xz}^{(\beta)}, \sigma_{yz}^{(\beta)} \sim r^{\lambda^{(\beta)}(T^*(z))-1} \sim r^{s^{(\beta)}(T^*(z))+1} \end{cases} \tag{13}$$

Substituting Eq. (13) into Eq. (1), the following strain rate exponents can be obtained,

$$\begin{cases} \dot{\varepsilon}_{xx}^{(\beta)}, \dot{\varepsilon}_{yy}^{(\beta)}, \dot{\varepsilon}_{xy}^{(\beta)}, \dot{\varepsilon}_{zz}^{(\beta)} \sim r^{n\left[\lambda^{(\beta)}(T^*(z))-2\right]} \sim r^{ns^{(\beta)}(T^*(z))} \\ \dot{\varepsilon}_{xz}^{(\beta)}, \dot{\varepsilon}_{yz}^{(\beta)} \sim r^{n\left[\lambda^{(\beta)}(T^*(z))-2\right]+1} \sim r^{ns^{(\beta)}(T^*(z))+1} \end{cases} \tag{14}$$

The detailed derivations of Eq. (12)~(14) have been listed in Appendix A. From Eq. (13) and Eq. (14), it is found that the in-plane components and the out-of-plane normal component of stress and strain rate are the leading terms near the notch tip. However, the out-of-plane shear components are of higher order which can be ignored in the asymptotic analysis. According to Eq. (9), the out-of-plane normal components $\sigma_{zz}$ and $\dot{\varepsilon}_{zz}$ can be determined by in-plane components. Hence, only the in-plane stress and strain rate components need to be considered in the asymptotic analysis. Considering Eq. (5), the in-plane components depend only on the Maxwell stress function $\chi_3$. As a result, only $\chi_3$ and in-plane compatibility equation are needed to construct the governing equation. The in-plane compatibility equation is written as follows:

$$\frac{\partial^2 \dot{\varepsilon}_{xx}}{\partial y^2} + \frac{\partial^2 \dot{\varepsilon}_{yy}}{\partial x^2} = \frac{\partial^2 \dot{\varepsilon}_{xy}}{\partial x \partial y} \tag{15}$$

Therefore, only the in-plane stress components and out-of-plane normal stress component are needed to establish the governing equation. The equation for the solution is simplified under the hypothesis mentioned above.

## 2.3 Hierarchy order asymptotic analysis

The asymptotic method is a mathematical method for series of expansion. It has successfully applied in solving the leading term of the tip field in 2D and 3D cases, such as the HRR solution (Hutchinson, 1968; Rice and Rosengren, 1968) and 3D tip field (Guo, 1993b). And for the higher-order term, the asymptotic method is also widely used during analysis, such as Xia et al. (1993), Yang et al. (1993b), Nguyen et al. (2000), as well as Loghin and Joseph (2020). The asymptotic analysis method is employed in the present paper which will be derived as the following.

### 2.3.1 Hierarchy order analysis

The following analysis in this paper is conducted in a cylindrical coordinate which is shown in Fig. 2b. Hence, the expansion form of the stress function $\chi_3$ is as follows:

$$\chi = \chi_3 = \sum_{\beta} A_{\beta}^{\mathrm{T}}(t) r^{s_{\beta}(T^*(z))+2} \tilde{\chi}^{(\beta)}(\theta, T_z) \qquad (16)$$

where $\tilde{\chi}^{(\beta)}(\theta, T_z)$ is the angular distribution function of stress function related to $T_z^*$. And $s_{\beta}(T^*(z))$ is the stress exponents depending on $T^*$, $s_{\beta}$ is assumed to obey:

$$s_1 < s_2 < s_3 < \cdots \qquad (17)$$

Herein, the stress exponents of the higher order term $s_{\beta}$ does not require that the higher order stress exponent should be also singular form although the stress exponent of the first order term is singular (referred to Riedel and Rice (1980) and Guo (1993a, 1993b)). Otherwise, the role of the leading term is always more significant than the second-order term based on the nature of asymptotic analysis. However, the fracture process zone size controlled by the leading term is generally relatively small. The finite deformation ahead of crack in real structures could also lower down the applicability of the leading term. The added higher order terms could enlarge the fracture parameter dominated region for tip fields. That is also one of the reasons to develop the higher order term solutions here.

The stress components are written as follows:

$$\begin{cases} \sigma_{rr} = \sum_{\beta} A_{\beta}^{\mathrm{T}}(t) r^{s_{\beta}(T^*(z))} \tilde{\sigma}_{rr}^{(\beta)}(\theta, T_z) \\ \sigma_{\theta\theta} = \sum_{\beta} A_{\beta}^{\mathrm{T}}(t) r^{s_{\beta}(T^*(z))} \tilde{\sigma}_{\theta\theta}^{(\beta)}(\theta, T_z) \\ \sigma_{r\theta} = \sum_{\beta} A_{\beta}^{\mathrm{T}}(t) r^{s_{\beta}(T^*(z))} \tilde{\sigma}_{r\theta}^{(\beta)}(\theta, T_z) \end{cases} \quad (18)$$

Combining Eq. (5) and Eq. (16), the relation between the angular distribution functions of stress function and stress components is obtained:

$$\begin{cases} \tilde{\sigma}_{rr}^{(\beta)}(\theta, T_z) = (s_{\beta}+2)\tilde{\chi}^{(\beta)} + \tilde{\chi}_{,\theta\theta}^{(\beta)} \\ \tilde{\sigma}_{\theta\theta}^{(\beta)}(\theta, T_z) = (s_{\beta}+2)(s_{\beta}+1)\tilde{\chi}^{(\beta)} \\ \tilde{\sigma}_{r\theta}^{(\beta)}(\theta, T_z) = -(s_{\beta}+1)\tilde{\chi}_{,\theta}^{(\beta)} \end{cases} \quad (19)$$

in which $()_{,\theta} = \partial()/\partial\theta$. It follows from Eq. (9) and Eq. (19) that

$$\sigma_{zz} = T_z \bullet \left[ \sum_{\beta} A_{\beta}^{\mathrm{T}}(t) r^{s_{\beta}(T^*(z))} \left( \tilde{\sigma}_{rr}^{(\beta)} + \tilde{\sigma}_{\theta\theta}^{(\beta)} \right) \right] \quad (20)$$

Hence, it is obtained that

$$\tilde{\sigma}_{zz}^{(\beta)} = T_z \bullet \left( \tilde{\sigma}_{rr}^{(\beta)} + \tilde{\sigma}_{\theta\theta}^{(\beta)} \right) \quad (21)$$

As discussed in Section 2.2, after ignoring $\sigma_{xz}$ and $\sigma_{yz}$, the deviatoric stress components are written as:

$$s_{ij} = \sum_{\beta} A_{\beta}^{\mathrm{T}} r^{s_{\beta}(T^*(z))} \tilde{s}_{ij}^{(\beta)}(\theta, T_z) \quad (22)$$

with

$$\begin{cases} \tilde{s}_{rr}^{(\beta)}(\theta, T_z) = \frac{2-T_z^*}{3}\tilde{\sigma}_{rr}^{(\beta)} - \frac{1+T_z}{3}\tilde{\sigma}_{\theta\theta}^{(\beta)} \\ \tilde{s}_{rr}^{(\beta)}(\theta, T_z) = \frac{2-T_z}{3}\tilde{\sigma}_{\theta\theta}^{(\beta)} - \frac{1+T_z}{3}\tilde{\sigma}_{rr}^{(\beta)} \\ \tilde{s}_{r\theta}^{(\beta)}(\theta, T_z) = \tilde{\sigma}_{r\theta}^{(\beta)} \\ \tilde{s}_{zz}^{(\beta)}(\theta, T_z) = \frac{2T_z-1}{3}\left( \tilde{\sigma}_{rr}^{(\beta)} + \tilde{\sigma}_{\theta\theta}^{(\beta)} \right) \end{cases} \quad (23)$$

According to Eq. (22) and Eq. (23), the expansion form of $\sigma_e^{n-1}$ is written as follows:

$$\sigma_e^{n-1} = \left(\frac{3}{2} s_{ij} s_{ij}\right)^{(n-1)/2}$$

$$= \left(A_1^T\right)^{n-1} r^{(n-1)s_1} \left(\tilde{\sigma}^{11}\right)^{n-1} \left\{1 + (n-1) \times \left[\frac{A_2^T}{A_1^T} r^{\Delta s_2} \tilde{\sigma}^{12} + \frac{A_2^T}{A_1^T} r^{\Delta s_3} \tilde{\sigma}^{13}\right.\right.$$

$$\left. + \cdots + \frac{1}{2}\left(\frac{A_2^T}{A_1^T}\right)^2 r^{2\Delta s_2} \tilde{\sigma}^{22} + \cdots\right]$$

$$\left. + \frac{(n-1)(n-3)}{2}\left[\left(\frac{A_2^T}{A_1^T}\right)^2 r^{2\Delta s_2} \left(\tilde{\sigma}^{12}\right)^2 + \cdots\right] + \cdots\right\} \qquad (24)$$

with

$$\begin{cases} \Delta s_q = s_q - s_1 \\ \tilde{\sigma}^{11}\left(\theta, T_z^*\right) = \left(\frac{3}{2} \tilde{s}_{ij}^{(1)} \tilde{s}_{ij}^{(1)}\right)^{1/2} \\ \tilde{\sigma}^{kq}\left(\theta, T_z^*\right) = \frac{3}{2} \frac{\tilde{s}_{ij}^{(k)} \tilde{s}_{ij}^{(q)}}{\left(\tilde{\sigma}^{11}\right)^2}; \; k+q \neq 2. \end{cases} \qquad (25)$$

In the following analysis, only the leading and the second-order terms are considered, as the combination of the leading and the second-order terms are sufficient to characterize the tip fields discussed below. Hence, $\sigma_e^{n-1}$ can be rewritten as:

$$\sigma_e^{n-1} = \left(\frac{3}{2} s_{ij} s_{ij}\right)^{(n-1)/2} = \left(A_1^T\right)^{n-1} r^{(n-1)s_1} \left(\tilde{\sigma}^{11}\right)^{n-1} \left(1 + (n-1) \times \frac{A_2^T}{A_1^T} r^{\Delta s_2} \tilde{\sigma}^{12} + \cdots\right) \qquad (26)$$

Substituting Eq. (26) into Eq. (1), the strain rate expansion containing the leading and the second-order terms is written as:

$$\dot{\varepsilon}_{ij} = \frac{\dot{\varepsilon}_0}{(\sigma_0)^n}\left(A_1^T\right)^n r^{ns_1} \tilde{\varepsilon}_{ij}^{(1)} + \frac{\dot{\varepsilon}_0}{(\sigma_0)^n}\left(A_1^T\right)^{n-1} A_2^T r^{ns_1+\Delta s_2} \tilde{\varepsilon}_{ij}^{(2)} + \cdots + A_1^T r^{s_1} \tilde{E}_{ij}^{(1)} + A_2^T r^{s_2} \tilde{E}_{ij}^{(2)} + \cdots$$

$$(27)$$

in which

$$\begin{cases} \tilde{\varepsilon}_{ij}^{(1)}\left(\theta,T_z^*\right) = \dfrac{3}{2}\left(\tilde{\sigma}^{11}\right)^{n-1}\tilde{s}_{ij}^{(1)} \\ \tilde{\varepsilon}_{ij}^{(m)}\left(\theta,T_z^*\right) = \dfrac{3}{2}\left(\tilde{\sigma}^{11}\right)^{n-1}\left[\tilde{s}_{ij}^{(m)} + (n-1)\tilde{\sigma}^{1m}\tilde{s}_{ij}^{(1)}\right];\ m = 2,\ 3,\ 4\cdots \\ \tilde{E}_{ij}^{(m)}\left(\theta,T_z^*\right) = \left(\dfrac{1+\nu}{E}\right)\tilde{s}_{ij}^{(m)} - \left(\dfrac{1-2\nu}{3E}\right)\tilde{\sigma}_{kk}^{(m)}\delta_{ij};\ m = 1,\ 2,\ 3\cdots \end{cases} \quad (28)$$

The compatibility relation equation in a cylindrical coordinate system is written as

$$\frac{1}{r}\left(r\dot{\varepsilon}_{\theta\theta}\right)_{,rr} + \frac{1}{r^2}\dot{\varepsilon}_{rr,\theta\theta} - \frac{1}{r}\dot{\varepsilon}_{rr,\,r} - \frac{2}{r^2}\left(r\dot{\varepsilon}_{r\theta,\theta}\right)_{,r} = 0 \quad (29)$$

in which $(\ )_{,r} = \partial(\ )/\partial r$. Substituting Eq. (27) and Eq. (28) into Eq. (29), the governing equation of asymptotic analysis is obtained:

$$\frac{\dot{\varepsilon}_0}{(\sigma_0)^n}\left(A_1^{\mathrm{T}}\right)^n r^{s_1}\cdot G_1\left[\tilde{\chi}^{(1)}(\theta,T_z);s_1\right] + \frac{\dot{\varepsilon}_0}{(\sigma_0)^n}\left(A_1^{\mathrm{T}}\right)^{n-1} A_2^{\mathrm{T}} r^{s_2}\cdot G_2\left[\tilde{\chi}^{(1)}(\theta,T_z),\tilde{\chi}^{(2)}(\theta,T_z);s_2\right]$$
$$+\cdots + A_1^{\mathrm{T}} r^{(2-n)s_1} G_{e1}\left[\tilde{\chi}^{(1)}(\theta,T_z)\right] + \cdots = 0$$

(30)

where

$$\begin{cases} G_1\left[\tilde{\chi}^{(1)}(\theta,T_z^*);s_1\right] = \left(\tilde{\varepsilon}_{rr}^{(1)}\right)_{,\theta\theta} - ns_1\left[\tilde{\varepsilon}_{rr}^{(1)} - (ns_1+1)\tilde{\varepsilon}_{\theta\theta}^{(1)}\right] - 2(ns_1+1)\left(\tilde{\varepsilon}_{r\theta}^{(1)}\right)_{,\theta} \\ G_2\left[\tilde{\chi}^{(1)}(\theta,T_z^*),\tilde{\chi}^{(2)}(\theta,T_z^*);s_2\right] \\ \quad = \left(\tilde{\varepsilon}_{rr}^{(2)}\right)_{,\theta\theta} - (ns_1+\Delta s_2)\left[\tilde{\varepsilon}_{rr}^{(2)} - (ns_1+1+\Delta s_2)\tilde{\varepsilon}_{\theta\theta}^{(2)}\right] - 2(ns_1+1+\Delta s_2)\left(\tilde{\varepsilon}_{r\theta}^{(2)}\right)_{,\theta} \\ G_{e1}\left[\tilde{\chi}^{(1)}(\theta,T_z^*)\right] = \left(\tilde{E}_{rr}^{(1)}\right)_{,\theta\theta} - s_1\left[\tilde{E}_{rr}^{(1)} - (s_1+1)\tilde{E}_{\theta\theta}^{(1)}\right] - 2(s_1+1)\left(\tilde{E}_{r\theta}^{(1)}\right)_{,\theta} \end{cases}$$

(31)

In Eq. (30) and Eq. (31), the equations of the first two terms in the asymptotic hierarchy are given. $G_1\left[\tilde{\chi}^{(1)}(\theta,T_z);s_1\right]$ corresponds to the leading term. $G_{e1}\left[\tilde{\chi}^{(1)}(\theta,T_z)\right]$ comes from the elastic part of the constitutive equations. The existence of $G_{e1}\left[\tilde{\chi}^{(1)}(\theta,T_z)\right]$ results in a classification discussion for the second term in the asymptotic hierarchy.

Furthermore, the boundary condition is needed to acquire the solution. For a sharp V-notch subjected to mode I loading, the symmetric condition at the symmetry plane ($\theta = 0$) and the traction-free condition at the notch flanks are expressed as:

$$\sigma_{rr,\theta}(\theta=0)=0, \sigma_{\theta\theta,\theta}(\theta=0)=0, \sigma_{r\theta}(\theta=0)=0 \qquad (32)$$

and

$$\sigma_{\theta\theta}(\theta=\pi-\alpha)=0, \sigma_{r\theta}(\theta=\pi-\alpha)=0 \qquad (33)$$

It follows from Eq. (5), Eq. (19), Eq. (32), and Eq. (33) that

$$\begin{cases} \tilde{\chi}^{(\beta)}_{,\theta\theta\theta}(0,T_z)=\tilde{\chi}^{(\beta)}_{,\theta}(0,T_z)=0 \\ \tilde{\chi}^{(\beta)}(\theta=\pm(\pi-\alpha),T_z)=\tilde{\chi}^{(\beta)}_{,\theta}(\theta=\pm(\pi-\alpha),T_z)=0 \end{cases} \qquad (34)$$

Hence, this problem is solvable as a fourth-order ordinary differential equation (ODE) with complete boundary conditions. The detailed solution process for solving the first two terms is given as following.

### 2.3.2 Leading term eigenvalue solutions

To obtain the leading term of the asymptotic solution, the first part in Eq. (30) need to be solved:

$$G_1\left[\tilde{\chi}^{(1)}(\theta,T_z); s_1\right]=0 \qquad (35)$$

The equation is a fourth-order ODE with respect to $\tilde{\chi}^{(1)}(\theta,T_z)$, and the boundary condition is given by Eq. (34) together with normalization $\tilde{\chi}^{(1)}(0,T_z)=1$. Hence, the boundary conditions can be expressed as follows:

$$\begin{cases} \tilde{\chi}^{(1)}(0,T_z)=1 \\ \tilde{\chi}^{(1)}_{,\theta}(0,T_z)=0, \\ \tilde{\chi}^{(1)}_{,\theta\theta}(0,T_z)=x \\ \tilde{\chi}^{(1)}_{,\theta\theta\theta}(0,T_z)=0 \end{cases} \qquad (36)$$

with

$$\begin{cases} \tilde{\chi}^{(1)}(\pm(\pi-\alpha),T_z)=0 \\ \tilde{\chi}^{(1)}_{,\theta}(\pm(\pi-\alpha),T_z)=0 \end{cases} \qquad (37)$$

An initial guess value $x$ is set in Eq. (36). The solution can be obtained by finding appropriate $s_1$ and $x$ to make $\tilde{\chi}^{(1)}(\theta,T_z)$ satisfy Eq. (37). The equation is solvable through the shooting method and a suitable iterative strategy. The stress exponents $s_1$ and the angular distribution function $\tilde{\chi}^{(1)}(\theta,T_z)$ affected by the out-of-plane effect are obtained through the solving process. The leading term has been solved and discussed systematically (Kong et al., 2022a). The results of the leading term are the basis of solving the second-order term.

**2.3.3 Second-order term eigenvalue solutions**

The solving process for the second-order term is divided into two steps according to the value of $s_2$. According to Eq. (30) and hierarchy order analysis, two conditions (cases) are considered here.

- Case I:

If $s_2 < (2-n)s_1$, the governing equation for the second-order term is written as:

$$G_2\left[\tilde{\chi}^{(1)}(\theta,T_z),\tilde{\chi}^{(2)}(\theta,T_z);s_2\right]=0, \text{ for } s_2<(2-n)s_1. \tag{38}$$

The boundary conditions are the same as Eq. (36) and Eq. (37) provided $\tilde{\chi}^{(1)}$ is replaced by $\tilde{\chi}^{(2)}$. Moreover, the solving strategy is also consistent.

- Case II:

On the condition of $s_2 = (2-n)s_1$, the governing equation turns into the following form:

$$\frac{\dot{\varepsilon}_0}{(\sigma_0)^n}\left(A_1^{\mathrm{T}}\right)^{n-1}A_2^{\mathrm{T}}r^{s_2}\cdot G_2\left[\tilde{\chi}^{(1)}(\theta,T_z),\tilde{\chi}^{(2)}(\theta,T_z);s_2\right]+A_1^{\mathrm{T}}r^{(2-n)s_1}G_{e1}\left[\tilde{\chi}^{(1)}(\theta,T_z)\right]=0 \tag{39}$$

To make the solution of angular distribution function independent of $A_2$, it is reasonable to set $A_2^{\mathrm{T}} = k\cdot\left(A_1^{\mathrm{T}}\right)^{2-n}\cdot(\sigma_0)^n/\dot{\varepsilon}_0$. Hence, Eq. (39) is expressed in the following form:

$$k\cdot G_2\left[\tilde{\chi}^{(1)}(\theta,T_z),\tilde{\chi}^{(2)}(\theta,T_z);s_2\right]+G_{e1}\left[\tilde{\chi}^{(1)}(\theta,T_z)\right]=0, \text{ for } s_2=(2-n)s_1 \tag{40}$$

The boundary conditions of Eq. (40) are the same as Eq. (36) and Eq. (37) if $\tilde{\chi}^{(1)}$ is replaced with $\tilde{\chi}^{(2)}$. However, the difference is that $s_2$ is a known quantity while $k$ is unknown in Eq. (40). Therefore, the solving target is finding appropriate $k$ and $x$ to satisfy the boundary condition. It is noted that the effects of elastic term enter the second-order term for Case II, hence the Poisson ratio is needed in solving the angular distribution function. Herein, the elastic term here is the requirement of asymptotic solution hierarchy structure which was originally induced by the leading term (Sharma and Aravas, 1991). The Poisson ratio 0.33 is used in this paper. However, the stress exponents are independent on the Poisson ratio if $s_2 = (2-n)s_1$. Similarly, a shooting method and iterative strategy are employed in the solving process.

The commercial mathematic solver MATLAB is employed in the solving process. A standard shooting numerical procedure and the fourth-order Runge-Kutta method are used. With appropriate initial guess variables, the solving process can be obtained within several seconds. Through solving the governing equations, the characteristic base quantities of the first two terms (i.e., $s_1$, $s_2$, $\tilde{\chi}^{(1)}(\theta, T_z)$, $\tilde{\chi}^{(2)}(\theta, T_z)$ ) are obtained.

## 3. Results and discussion

The characteristic parameters of tip fields for 3D sharp V-notch subjected to mode I creeping loading are given through asymptotic analysis. In this section, a novel tip field considering 3D constraint effects is established, and FEA analysis is executed for verification.

### 3.1 3D sharp V-notch tip field with constraint effect

The first two order terms of tip fields are considered. Hence, the stress expansion is written as:

$$\frac{\sigma_{ij}}{\sigma_0} = K_N^T(t, T_z) r^{s_1(T^*(z))} \hat{\sigma}_{ij}^{(1)}(\theta, T_z) + A_2^T(t, T_z) r^{s_2(T^*(z))} \hat{\sigma}_{ij}^{(2)}(\theta, T_z) \qquad (41)$$

in which $K_N^T(t, T_z)$ denotes the notch stress intensity factor dependent on creep time and out-of-plane factor. $A_2^T(t, T_z)$ represents the amplitude of the second-order term which is also dependent on creep time and out-of-plane factor. $s_1$ and $s_2$ represent the stress exponents of the leading and second-order terms, respectively. Moreover, the $\hat{\sigma}_{ij}^{(1)}$ and $\hat{\sigma}_{ij}^{(2)}$ are the stress angular distribution functions of the first two terms. The stress expansions Eq. (41) and Eq. (18) have the same meaning. However, different normalization strategies are used. In Eq. (18), the determination of amplitude and angular distribution functions is based on the normalization condition $\tilde{\chi}^{(\beta)}(0,T_z)=1$. While in Eq. (41), the normalization strategies are set as $\left(\hat{\sigma}_e^{(1)}\right)_{max}=1$ and $\hat{\sigma}_{\theta\theta}^{(2)}(\theta=0)=1$.

A novel tip field solution is defined in Eq. (41), i.e., $K_N^T(t,T_z)-A_2^T(t,T_z)$ solution. The proposed solution considers the out-of-plane effect characterized by $T_z$ in 3D fracture problems. Moreover, a constraint factor $A_2^T$ is proposed to describe the second-order term in 3D tip fields. It should be noted that not only the leading term but also the higher-order term is affected by the out-of-plane effect according to Eq. (41). In other words, the $T_z$ in parentheses $K_N^T(t,T_z)-A_2^T(t,T_z)$ not only means that the amplitude is affected by $T_z$, but also that the characteristics of $K_N^T$-term and $A_2^T$-term varies with $T_z$, including the stress exponents and angular distribution functions. For the convenience of expression, it is directly expressed as $K_N^T-A_2^T$ solution in the following discussion.

The proposed $K_N^T-A_2^T$ solution demonstrates the effect of the 3D constraint completely and explicitly in theoretical form. The 3D constraint is contributed by both in-plane and out-of-plane effects. However, since the $A_2^T$ term is also influenced by $T_z$, the in-plane and out-of-plane effects on constraint level are highly interlinked and interacted. The specific explanation and discussion are shown below. The effect of out-of-plane effect on the leading term of notch tip fields has been reported in our previous study (Kong et al., 2022a). Hence, the following results are mainly about the second-order term.

### 3.1.1 Stress exponents of the second-order term

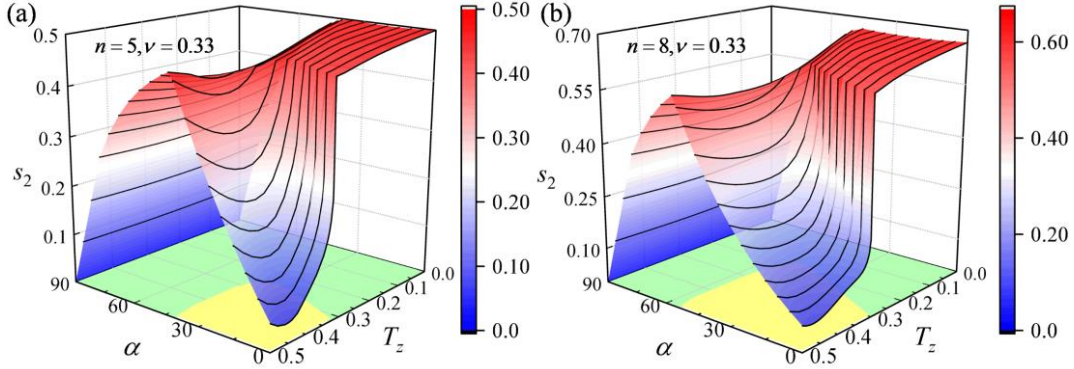

Fig. 3 Stress exponents of the second-order term for notches with different angles with (a) $n = 5$, $\nu = 0.33$ (b) $n = 8$, $\nu = 0.33$

As shown in Fig. 3, the stress exponents of the second-order term for notches with different angles are presented. It is found that $s_2$ is significantly affected by out-of-plane factor $T_z$ and notch angle $2\alpha$. When $\alpha$ is $90°$, the second-order stress exponent is equal to 0. The yellow projection of the surface on $\alpha - T_z$ plane represents that $s_2$ is determined from Eq. (38), i.e., $s_2 < (2-n)s_1$ (Case I ). While the green projection (see Fig. 3) represents that $s_2$ is identical to $(2-n)s_1$ (Case II). By comparing Fig. 3a and b, the proportion of Case I increases with creep exponent $n$. Moreover, Case I happens when $\alpha$ approaches 0 and $T_z$ approaches 0.5. In other words, $s_2 < (2-n)s_1$ happens with small notch angles and large $T_z$. However, it should be noted that the effect of $T_z$ is slight when $\alpha$ is sufficiently large for Case II. For $n = 5$ and $n = 8$, the second-order stress exponents $s_2$ is always larger than or equal to 0, which indicates that the second-order term has no singularity.

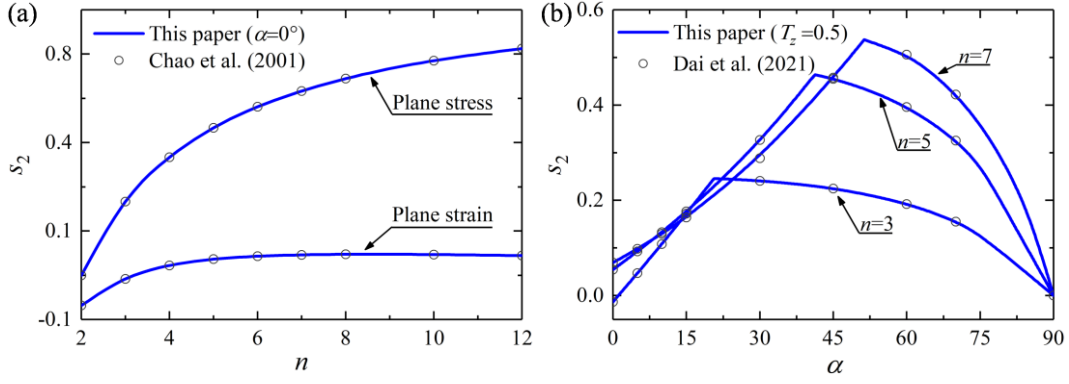

Fig. 4 Comparison of the present paper and previous studies: the second-order stress exponents of (a) cracks (b) notches under plane strain condition

The second-order stress exponents of 3D notches can be degenerated to the 2D crack cases. When $\alpha$ is 0°, the notch is equivalent to an ideal crack. According to Section 2.2, the 3D stress exponents degenerate to plane-stress condition with $T_z = 0$. The problem of how to degenerate the 3D second-order term to the-plane strain solution needs to be discussed separately. For Case I in which $s_2 < (2-n)s_1$, the solution is not affected by the elasticity. Hence, the creep predominant condition is kept. The plane-strain condition $\varepsilon_{zz} = 0$ requires $T_z = 0.5$. For Case II in which $s_2 = (2-n)s_1$, $T_z = 0.5$ is no longer a necessary condition for plane-strain state as elasticity effect enters the second-order term. However, the stress exponents are not affected by the elasticity as $s_2 = (2-n)s_1$ is kept for Case II. It should be noted that plane-strain state is a 3D stress state in which $T_z$ transits from 0.5 to the Poisson ratio at the creep border. The 3D asymptotic solutions can be employed to describe plane-strain tip field by introducing the effect of $T_z$.

As shown in Fig. 4a, the second-order stress exponents of cracks ($\alpha = 0°$ in the present paper) agree well with the results in the previous article (Chao et al., 2001). In addition, the present $s_2$ of 3D notches degenerates to plane-strain cases. As shown in Fig. 4b, the present results are in good agreement with $s_2$ of notches under plane-strain condition (Dai et al., 2021). The comparison shows that the present results can be degenerated to particular cases (e.g., cracks and notches under 2D plane conditions) and show good accuracy.

**3.1.2 Stress distribution function of the second-order term**

In addition to the second-order stress exponents, the stress angular distribution functions are also obtained. The second-order angular distribution functions with various $T_z$ for $2\alpha = 30°$ and $2\alpha = 120°$ are presented in Figs. 5~6, respectively. The colors of the projection in $[\theta/(\pi-\alpha)]-T_z$ plane have the same meaning as those already shown in Fig. 3. It is concluded that for notches with $30°$ angle, the stress angular distribution functions are determined from Eq. (38) (Case I, yellow projection) when $T_z$ is larger than 0.29 for $n = 5$. However, when the notch angle becomes large, the stress angular distribution functions are totally determined from Eq. (40) (Case II, green projection). Fig. 5 shows that the angular distribution functions are significantly affected by the out-of-plane factor $T_z$. When $T_z$ approaches 0, the angular distribution function changes dramatically, especially near the notch flanks (i.e., $\theta/(\pi-\alpha) \to 1$). The phenomenon is also manifested in the leading term of the tip fields (Kong et al., 2022a). For notches with small angles, the more closely the 3D stress state approaches the plane-stress state, the more complicated the variation of the angular distribution function is. In the previous literature (Loghin and Joseph, 2020; Yang et al., 1993a), the angular distribution functions of cracks under plane-stress conditions show the same and universal complexity. Moreover, for notches with large angles (e.g., $2\alpha = 120°$, see Fig. 6), the second-order stress angular distribution is still seriously dependent on $T_z$, which is different from that of the leading term. For large angle notches, the effect of $T_z$ on the angular distribution functions of the leading term is limited (Kong et al., 2022a). While for the second-order angular distribution functions, only $\hat{\sigma}_{\theta\theta}^{(2)}$ is slightly affected by $T_z$.

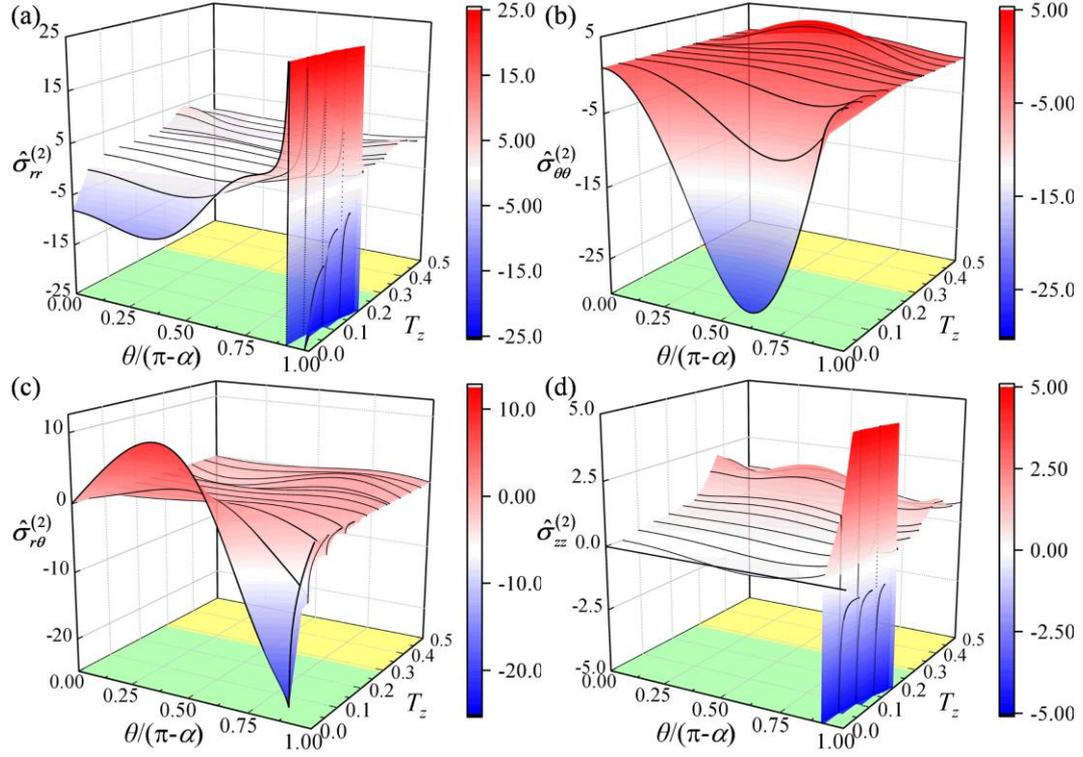

Fig. 5 The second-order stress angular distribution function for notches under various $T_z$ when $2\alpha = 30°$ and $n = 5$, $\nu = 0.33$

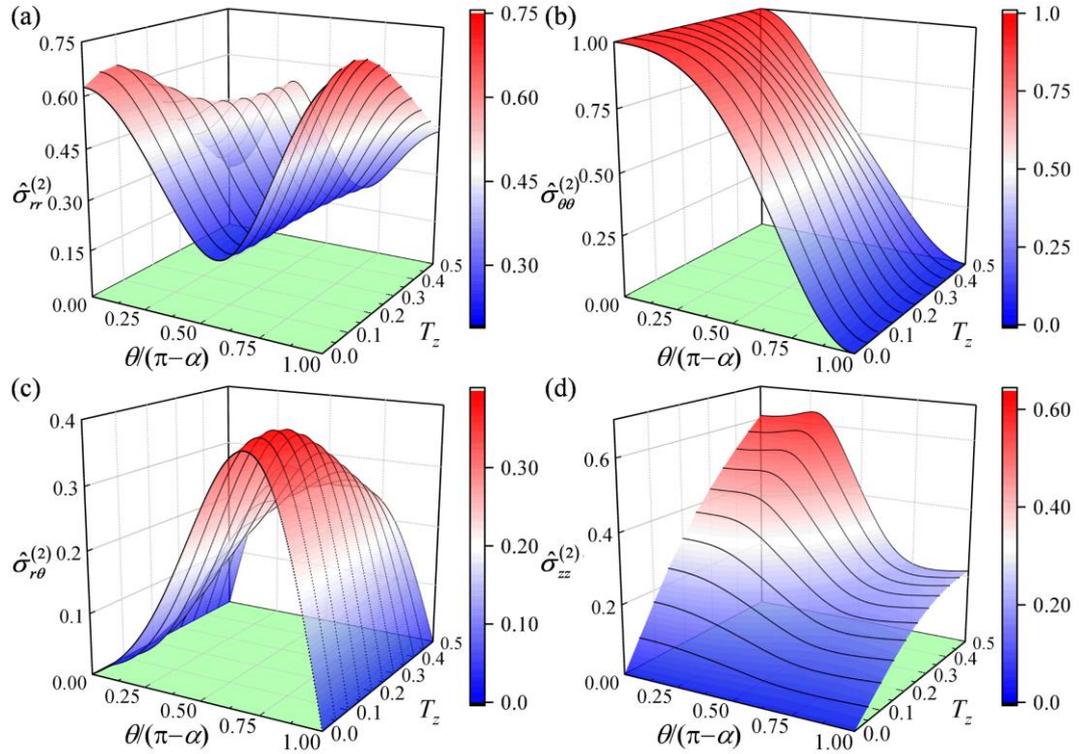

Fig. 6 The second-order stress angular distribution function for notches under various $T_z$ when $2\alpha = 120°$ and $n = 5$, $\nu = 0.33$

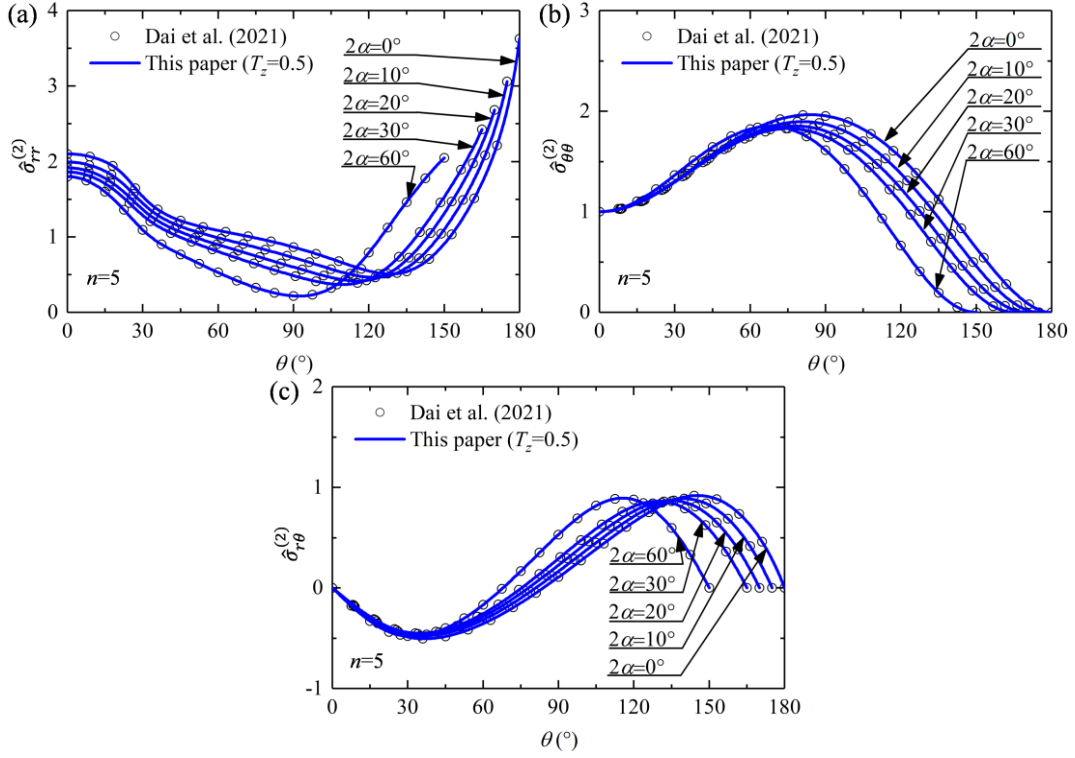

Fig. 7 The comparison of second-order stress angular distribution function under plane-strain condition

The stress angular distribution function can also be degenerated to 2D plane condition. To the best of the authors' knowledge, there is no relative research giving the second-order stress angular distribution function under plane-stress condition. Hence, the results under plane-strain are compared only. For plane-strain condition, $T_z$ transits from 0.5 to Poisson ratio (0.33 in this paper) at the creep border. In previous literature, the creep predominant condition is assumed. Hence, when $T_z$ is identical to 0.5, the present second-order stress angular distribution function is in accord with that of 2D plane-strain cases in the previous paper (Dai et al., 2021). It is noted that the cases presented in Fig. 7 obey the condition of Case I in which the elasticity has no effect. Hence, it shows good agreement with the plane-strain solution deduced based on creep predominant assumption.

In general, the 3D second-order stress exponents and angular distribution functions are closely related to the out-of-plane factor $T_z$. Hence, for a 3D sharp V-notch, it is necessary to consider $T_z$ in the analysis of higher-order terms. $T_z$

characterizes the level of out-of-plane effect. The 3D stress exponents $s_\beta$ and angular distribution function $\tilde{\sigma}_{ij}^{(\beta)}$ can degrade into 2D plane cases when $T_z$ takes the appropriate value. The relative solution can describe the 3D stress state between plane-strain and plane-stress conditions. Moreover, the solution gives the angular distribution functions of the out-of-plane normal stress $\sigma_{zz}$ which has not attracted sufficient attention in 2D plane solutions. Much more interesting, it is common for the second-order governing equation to be affected by the elastic part of the constitutive equation (i.e., Case II, see Eq. (40)) after introducing the out-of-plane factor $T_z$. This characteristic of 3D notches or cracks differs from that of 2D cases. For 2D plane-strain cases, the second-order governing equation will not be affected by the elastic part of the constitutive equation (Xia et al., 1993).

## 3.2 Verification and comparison

### 3.2.1 The analysis of 3D notches in thick plate

A 3D notched thick plate is adopted as the finite element model. The detail FE model is given in Appendix B. A normalized creep time is employed as a reference in the following analysis and discussion (Zhu et al., 2001):

$$\tau = \frac{B\sigma_g^n t}{\sigma_g / E} \tag{42}$$

Fig. 8 presents the radial distribution of circumferential stress $\sigma_{\theta\theta}$ along the angular bisector direction of the notches with $2\alpha = 30°$ and $2B/a = 3$. Fig. 8a and Fig. 8b show the stress distribution at $\tau = 0.3323$ with $z/B = 0$ and $z/B = 2/3$, respectively. Fig. 8c and Fig. 8d show the stress distribution at $\tau = 0.0341$ with $z/B = 0$ and $z/B = 2/3$, respectively. According to Fig. 8, the FEA results have good agreement with the plane-strain (PE) leading term near the notch tip as the specimen is relatively thick. Despite that, the agreement with FEM results diminishes as the radial distance ($r/a$) increases due to the fact that the area away from the notch tip is elastic-dominated rather than creep-dominated. Then, the plane-strain two-term solution considering the in-plane constraint has the advantage of agreeing with the FEA results. In addition, the 3D leading term considering the out-of-plane effect shows advantages in the distance from the tip, especially in the tendency of stress drop. The phenomenon indicates that the out-of-plane effect has a more apparent influence than the in-plane constraint in 3D sharp V-notch cases, especially in the plane near the free surface ($z/B = 2/3$). However, the 3D two-term solution ($K_N^T - A_2^T$ solution) considering 3D constraint effect has significant progress in describing the stress distribution of 3D sharp V-notch compared with all the previous theoretical solutions, especially when the distance from the notch tip is far.

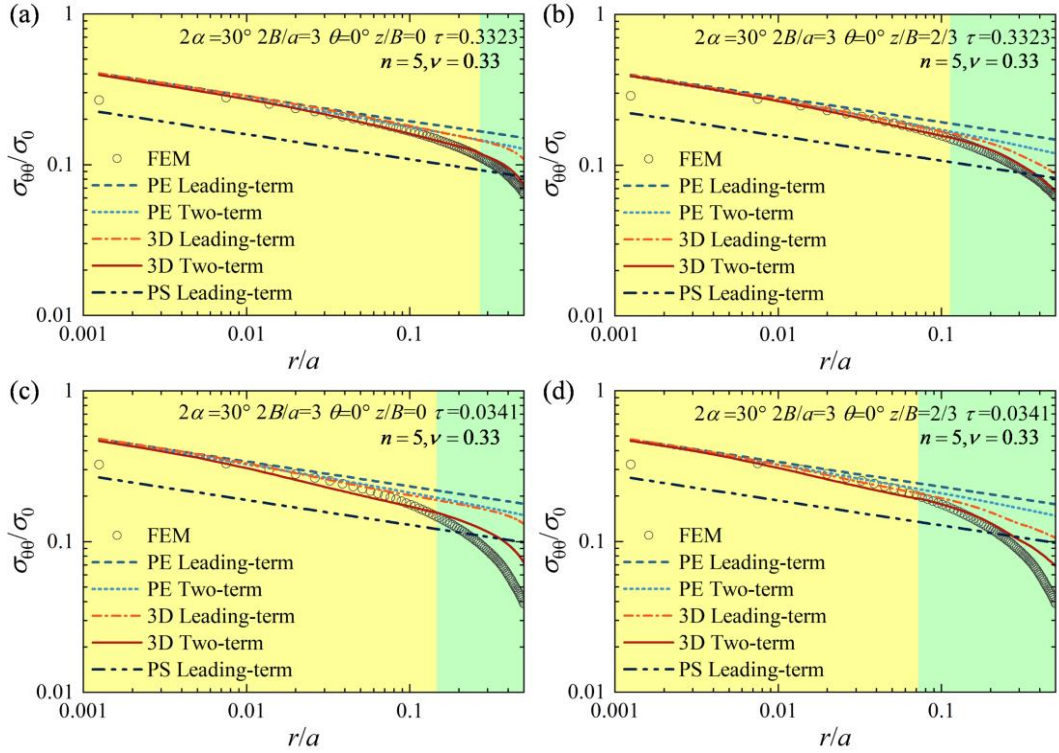

Fig. 8 The radial distribution of circumferential stress of 3D notch with 30° opening angle at different creep times and locations of the thick specimen ($2B/a = 3$)

It is found that the plane-stress (PS) solution cannot show good agreement with FEA results for thick specimens. Thus, the second-order term prediction based on plane-stress theory for thick specimen is not given in Fig. 8. It has been also given in Section 3.1 that $s_2 > 0$ for $n = 5$. Hence, the second-order term approach 0 when $r$ approaches 0 under plane stress condition. Accordingly, the second-order term based on plane-stress theory for the 3D thick specimen is meaningless.

Furthermore, the yellow (Case I) and green (Case II) backgrounds in the figures have the same meaning as the projection color of the stress exponents and angular distribution functions figures. Then, it represents that Case I and II co-exist in one 3D sharp V-notch specimen as the variation of the out-of-plane factor $T_z$. In other words, for a 3D sharp V-notch, the second-order term is affected by the elastic part of the constitutive relation (Case II) at the location away from the notch tip. While in the 2D plane cases, the co-existence of Case I and Case II has not been found yet (Xia et al., 1993). More specifically, the area dominated by Case I (yellow part) expands with increasing creep time. However, the area dominated by Case II (green part) shows its vitality when the creep time is not sufficiently long (see Fig. 8c, d), especially in the plane away from the middle plane of the specimen (see Fig. 8d). The creep-dominated area is surrounded by the elastic-dominated area. The elastic-dominated area remarkably influences the tip field of the sharp V-notch when the creep time is not very long (Kong et al., 2022a). Hence, the relative error between the 3D two-term solution ($K_N^T - A_2^T$ solution) and the FEA results is pronounced, as shown in Fig. 8c, d. Despite this, $K_N^T - A_2^T$ solution still has the best agreement with the FEA results compared with other theoretical solutions.

**3.2.2 The analysis of 3D notches in thin plates**

Fig. 9 presents the radial distribution of $\sigma_{\theta\theta}$ for 3D notch with 30° opening angle in the thin specimen ($2B/a = 0.1$). The characteristics of different theoretical solutions are clarified with the example of 3D thin sharp V-notch specimens. One can find that the FEA results tend to agree with the plane-strain solution when approaching the notch tip. The FEA results could agree with the plane-stress solution with the increase of radial distance. The characteristic of 3D notches in thin plates is similar to that of 3D cracks in thin plates (Yi and Wang, 2020). The plane strain two-term solution shows its advantage in agreeing with the transition of FEA results from plane-strain to plane-stress condition. However, the plane strain two-term solution is not accurate enough with increasing radial distance when the stress state in the thin specimen is under the control of plane-stress condition. This is reasonable as the plane-strain high-order term solution cannot describe the stress state under plane-stress condition.

On the other hand, the leading-term and the two-term solutions based on plane-stress conditions agree well with the FEA results when the radial distance gets large. And the plane-stress two-term solution has a slight advantage over the leading-term solution in accuracy with the FEA results when $r/a > 0.05$ (see Fig. 9a, b). Meanwhile, it is found that the 3D solutions considering the out-of-plane effect, including 3D leading-term and two-term solutions, have better performance in terms of agreement with FEA results. The 3D solutions have the capability of describing the transition from plane-strain to plane-stress condition. And the 3D two-term solution ($K_N^T - A_2^T$ solution) has better accuracy with increasing radial distance than the leading-term solution. However, the effect of the 3D second-order term in $K_N^T - A_2^T$ solution on the tip fields for thin plates is less pronounced than that for thick specimens, as shown in Fig. 8a, b. It indicates that the effect of the 3D constraint is related to the thickness of the specimen. In other words, the out-of-plane effect influences the 3D constraint, as discussed in Section 3.1. Moreover, it is concluded that the 3D second-order term is much more easily affected by the elastic part of the constitutive relation (Case II) near the notch tip by comparing the green area in Fig. 9 with that in Fig. 8.

Undeniably, the 3D leading-term and two-term solutions considering the out-of-plane effect have an obvious advantage for 3D sharp V-notch in thin plates compared with the 2D plane solutions. The reason is that the 3D solution can quantify both the plane-strain state and the plane-stress state, as well as the transition from plane-strain to plane-stress state in the thin specimen near the notch front. The second-order term solved based on the 3D constraint theory (i.e., $A_2^T$-term) is much more reasonable and practical than the second-order term based on plane-strain condition.

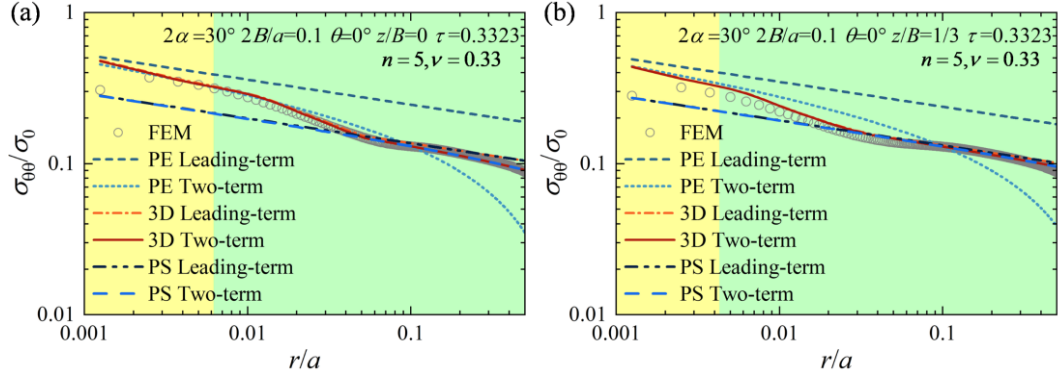

Fig. 9 The radial distribution of circumferential stress of 3D notch with 30° opening angle at different locations of the thin specimen ($2B/a = 0.1$)

### 3.2.3 The stress fields of 3D notches with large angles

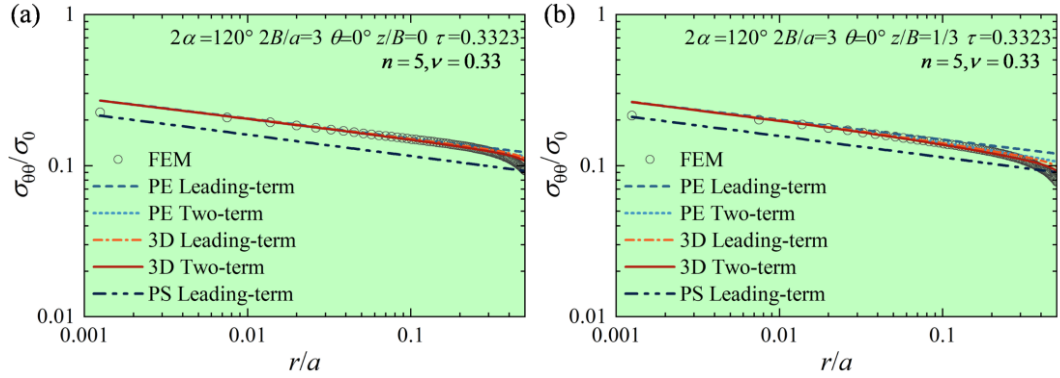

Fig. 10 The radial distribution of circumferential stress of 3D notch with 120° opening angle at different locations

To verify the validity of $K_N^T - A_2^T$ solution on a 3D sharp V-notch with large opening angle, the FE analysis of the 3D notch with 120° opening angle is conducted. Fig. 10 presents the radial distribution of circumferential stress $\sigma_{\theta\theta}$ of the 3D notch with 120° opening angle. It is found that the stress fields near the stress are totally affected by the elastic part of the constitutive relation (Case II) for $2\alpha = 120°$. This can also be concluded according to Fig. 3. The stress fields situation belongs to Case II regardless of the value of $T_z$ for $\alpha = 60°$ and $n = 5$. In addition, the 3D solutions considering the out-of-plane effect show better agreement with FEA results than the 2D solutions. Similar to those of the notches with opening angle 30°, the two-term solutions have better agreement with FEA results, especially the 3D two-term solution.

However, there is only a slight advantage over the 3D leading-term solution. Hence, the 3D constraint effect has less influence on the tip fields for notches with large opening angles. Hence, the 3D leading-term solution is sufficient to describe the stress field for notches with large notch opening angles. Adding of the second-order term does not significantly improve the prediction accuracy of the 3D notch tip stress fields for large notch specimens regardless of thick or thin plates.

### 3.2.4 The advantages of $K_N^T - A_2^T$ solution in stress angular distribution

In order to have a better understanding of the 3D constraint effect, the FEA results and theoretical solutions of the angular distribution of the tip fields are given. Fig. 11 shows the circumferential distribution of stress components of 3D notches in the thick specimen ($2B/a = 3$) with 30° opening angles. The results show that the 3D two-term solution agrees better with the FEA results than the 3D leading-term solution. It indicates that the $K_N^T - A_2^T$ solution considering the 3D constraint effect is a more accurate tip field solution close to the truth. In addition, the $K_N^T - A_2^T$ solution gives the distribution of out-of-plane normal stress $\sigma_{zz}$ which cannot be obtained from the 2D plane solutions. However, it should be noted that the $K_N^T - A_2^T$ is less accurate when the creep time is not long enough, as shown in Fig. 11b. The reason is that the elastic-dominated area affects the tip fields when the creep time is not sufficiently long.

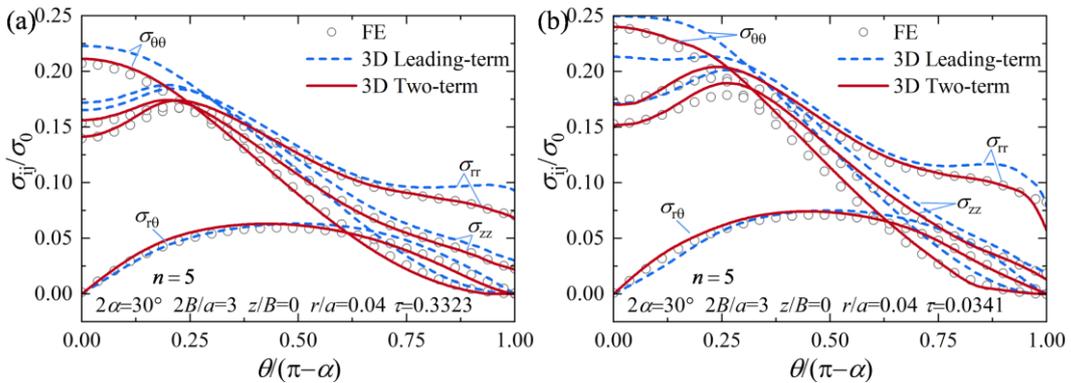

Fig. 11 Circumferential distribution of stress components in the 30° thick notch specimen at (a) $\tau = 0.3323$ (b) $\tau = 0.0341$

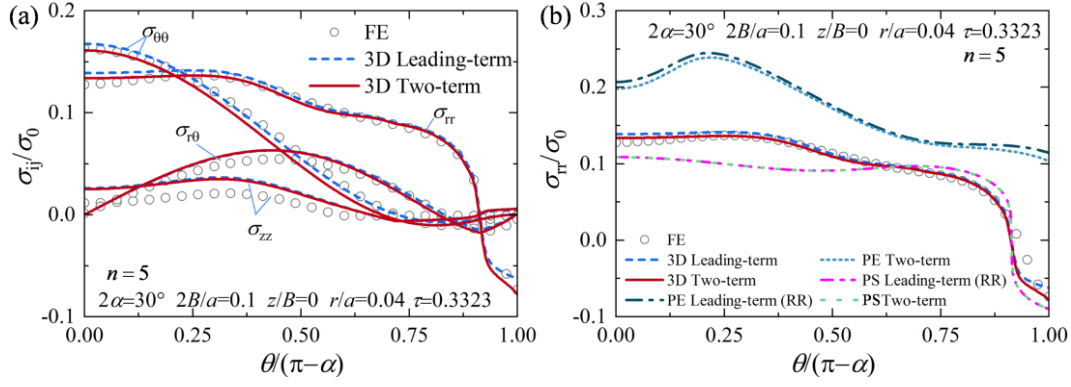

Fig. 12 (a) Circumferential distribution of stress components (b) comparison of 3D solutions and 2D solutions in the 30° thin notch specimen

As discussed above, the 3D two-term solution has a slight advantage over the 3D leading-term solution in describing the 3D tip fields for the notches in thin plates. This can also be concluded from the comparison of the circumferential distribution of stress components shown in Fig. 12a. This further proves that the 3D constraint effect is limited for the 3D sharp V-notches in thin plates. However, the slight advantage does not mean that the 3D constraint for 3D notches in thin plates is meaningless. As shown in Fig. 12b, only the 3D two-term solution obtained based on the 3D constraint theory is valid for the 3D notches. In contrast, the two-term solutions under plane-strain or plane-stress conditions are out of operation. In a word, the second-order term must be deduced based on the 3D constraint theory considering the out-of-plane effect for 3D notches to get a reasonable and accurate solution. Moreover, for notches in thin plates, the 3D leading-term solution considering $T_z$ shows good accuracy for stress field prediction.

Generally, the 3D two-term solution, i.e., $K_N^T - A_2^T$ solution deduced from the 3D constraint theory considering the out-of-plane effect has good accuracy in describing the tip fields of the 3D sharp V-notch. The $K_N^T - A_2^T$ solution has the advantages of higher-order solutions considering the constraint effect and 3D solutions considering the out-of-plane effect. Only the $K_N^T - A_2^T$ solution based on the 3D constraint effect gives a reasonable and accurate description of the tip fields of 3D sharp V-notches.

## 3.3 The distribution of stress exponents along the thickness

The distributions of stress exponents along the thickness at the notch front are obtained based on the $K_N^T - A_2^T$ solution. Fig. 13 shows the stress exponents of the leading and second-order terms along the thickness direction for sharp V-notch with $2\alpha = 30°$ and $2B/a = 3$. The out-of-plane factor $T_z$ is obtained from the FEM. The out-of-plane factor $T_z$ is found to keep close to 0.5 along the notch front near the middle plane and rapidly reduce to 0 approaching free surfaces. Combining with the results already shown in Fig. 3, the corresponding stress exponents are obtained. It indicates that the stress state near the middle plane of the thick specimen approaches the plane-strain condition. In a relatively large range, the first-order and second-order stress exponents are closer to those of plane-strain for thick specimen. Moreover, the first and second-order stress exponents change sharply near the free surface, indicating the complex stress state near the free surface of the 3D sharp V-notch.

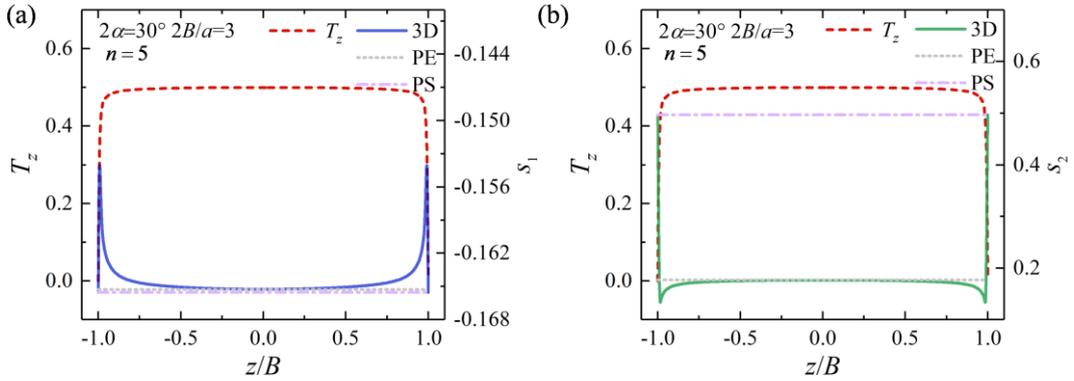

Fig. 13 The distributions of stress exponents and $T_z$ along the thickness at the notch front in thick specimen (a) first-order (b) second-order

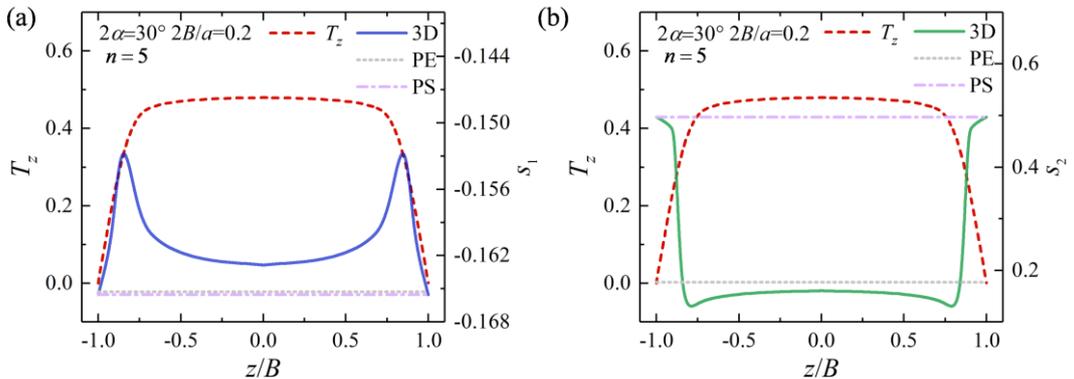

Fig. 14 The distributions of stress exponents and $T_z$ along the thickness at the notch front in thin specimen (a) first-order (b) second-order

For comparison, the distribution of $T_z$ and stress exponents throughout the thickness for thin specimen are shown in Fig. 14. A stress state close to plane-strain condition exists at the mid-plane even for the 3D notch in thin plates. The phenomenon can also be found in Fig. 9 that there is a transition opening stress from plane-strain condition to plane-stress condition along the radial direction. The existence of plane-strain condition at the mid-plane of the 3D cracks in thin plates is also discussed by Nakamura and Parks (1990) and Yi and Wang (2020). However, it should be noted that there is a slight difference between the stress state at the mid-plane and the plane-strain condition. As shown in Fig. 14, there is apparent difference of the 3D stress exponents from the plane-strain stress exponents in the thin plate. $T_z$ and stress exponents of 3D notches in thin plates change less rapidly than those in thick plates near the free surface. Similarly, the stress state turns into the plane-stress condition approaching the free surface.

## 4. 3D crack tip fields

As discussed in Section 3.1, the sharp V-notch is equivalent to an ideal crack when $\alpha$ is 0°. And the stress exponents for crack ($\alpha = 0°$) can also be obtained from the governing equation. The second-order stress exponents considering the out-of-plane effect for cracks are already shown in Fig. 3. The results are proven to be able to degenerate to the 2D plane crack cases (see Fig. 4a). For cracks, the 3D leading term is often characterized by the $C^{\mathrm{T}}(t, T_z)$-integral:

$$C^{\mathrm{T}}(t,T_z) = \int_{\Gamma}\left(\frac{n}{n+1}\sigma_{ij}\dot{\varepsilon}_{ij}n_1 - \sigma_{ij}n_j\dot{u}_{i,1}\right)ds \tag{44}$$

in which the stress, strain and displacement components are related to the out-of-plane effect. $\Gamma$ is a vanishingly small contour surrounding the crack tip. $n_i$ is the outer normal unit vector of $\Gamma$. $\dot{u}_{i,1}$ represents the displacement gradient rate.

Hence, the relation between the $C^T(t,T_z)$ and $K_N^T(t,T_z)$ is shown below:

$$K_N^T(t,T_z;\alpha=0)=\left(\frac{C^T(t,T_z)}{\sigma_0\dot{\varepsilon}_0 I_n(T_z,n)L}\right)^{-s_1} \tag{45}$$

in which $L$ is the characteristic length, and $I_n(T_z,n)$ is an integration constant as a function of creep exponents $n$ and out-of-plane factor $T_z$. The detailed analysis about $I_n(T_z,n)$ has been presented by Xiang and Guo (2013).

Accordingly, a higher-order solution for 3D cracks, i.e., $C^T(t,T_z)-A_2^T$ solution, can be obtained from the degeneration of $K_N^T-A_2^T$ solution. It is noted that the $C^T(t,T_z)-A_2^T$ solution is proposed for 3D crack tip fields in which the out-of-plane effect and the 3D constraint effect are taken into consideration. To verify the accuracy of the $C^T(t,T_z)-A_2^T$ solution for 3D cracks, the FE analyses are conducted. The geometry of the 3D crack is set as $2B/a=3$ and $a/b=1$. And the angle $2\alpha$ is identical to $0°$ for cracks. The comparisons of the asymptotic solutions and FEA results are shown in Fig. 15 and Fig. 17.

Fig. 15 presents the radial distribution of $\sigma_{\theta\theta}$ for 3D cracks. Similar to the 3D notches with small opening angles, it is found that the 3D two-term solution ($C^T(t,T_z)-A_2^T$ solution) has the best accuracy with the FEA results. The plane-strain leading-term and the two-term solutions have good agreement with the FEA results when the radial distance from the notch tip is small. In other words, the plane-strain solutions can describe the stress fields for the 3D crack near the notch tip as the specimen is relatively thick. The relative error gets pronounced with increasing radial distance without considering the out-of-plane effect for 3D cases. And the performance of the 3D leading-term solution is not satisfactory for deep cracks ($a/b=1$ in this example, see Fig. 15) when the radial distance is considerable. For thick specimen in which the stress state is close to plane strain condition, the 2D three-term solution can quantify the stress field better compared with the plane-strain two-term solution. And the performance of plane-strain three-term solution ($C(t)-A_2$) may be better than 3D leading-term within a certain range as shown in Fig. 15. However, when the radial distance gets large, the 3D two-term solution still have better agreement with the FE results. Hence, the 3D constraint effects must be considered for 3D specimens with deep cracks. Moreover, it is noted that for 3D crack, the tip field also contains both Case I and Case II areas in one specimen as $T_z$ varies for 3D conditions.

The 2D three-term solution is a significant solution in quantifying the tip fields in 2D cases, especially for plane-strain condition (Yang et al., 1993b; Zhu and Chao, 1999). As shown in Fig. 15, the plane-strain three-term solution has better agreement than the plane-strain two-term solution. The stress state in the mid-plane of the thick specimen is close to the plane-strain state. Hence, the plane-strain three-term solution have good agreement with the FE results. To clarify the effectivity of 2D three-term solution in detail, the radial distribution of circumferential stress of the 3D crack in thin specimen is shown in Fig. 16. Similar to 3D notches in thin specimen, the relative error between the 3D leading-term and the two-term solution is small. As discussed by Nakamura and Parks (1990) and Yi and Wang (2020), the stress state near the crack front in the mid-plane is close to the plane-strain condition and will turn into plane-stress condition with increasing radial distance. The phenomenon is also found from Fig. 16 and decide the effectivity of the higher-order term for 3D cracks in thin specimen. The plane-stress solution is only effective when the radial distance is large enough and the stress state is under plane-stress condition. Moreover, the plane-strain two-term solution and three-term solution have good agreement with FE results near the notch front when the stress state is under plane-strain condition. And the plane-strain three-term solution is more accurate than the plane-strain two-term solution when the stress state is under the transition from plane-strain to plane-stress state. Moreover, the 3D leading-term and 3D two-term solution have good agreement with the FE results for 3D cracks in thin specimen (see Fig. 16). The reason is that the 3D leading-term and 3D two-term solution considering the effect of $T_z$ can quantify the stress field better when the stress state changes from plane-strain to plane-stress state. Hence, for 3D cracks in thick specimen, the three-term solution is considerably effective. And the 3D solution considering $T_z$ is relatively accurate for 3D cracks in thin specimen.

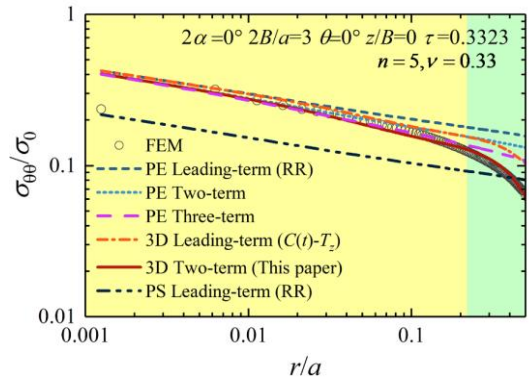

Fig. 15 The radial distribution of circumferential stress of the 3D crack in thick specimen at

$\tau = 0.3323$

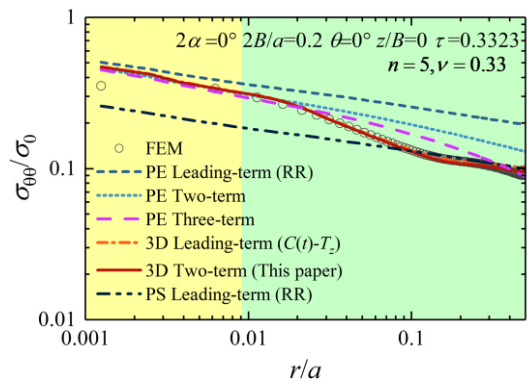

Fig. 16 The radial distribution of circumferential stress of the 3D crack in thin specimen at

$\tau = 0.3323$

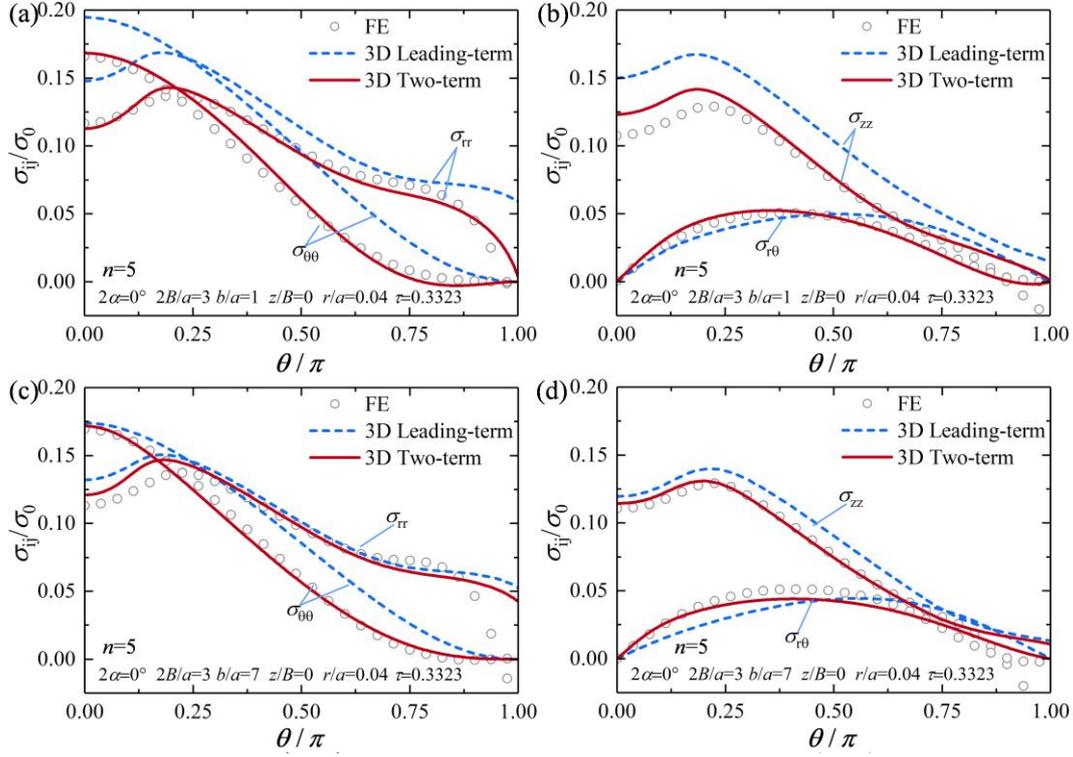

Fig. 17 Circumferential distribution of stress components in the 3D crack specimen with different relative crack depth

Furthermore, the circumferential distributions of stress components are also shown in Fig. 17 to present the effectivity of the 3D two-term solution. As expected, the 3D two-term solution has better agreement with the FEA results than the 3D leading-term solution for the 3D deep crack. In fact, as shown in Fig. 17a and b, the relative error between the 3D leading-term solution and FEA results is so pronounced that the 3D leading-term solution cannot describe the stress distribution for 3D deep cracks ($b/a=1$). Herein, it is emphasized that the 3D constraint effect is of great significance for 3D crack front, which can be considered as the worst case of the sharp V-notch with respect to the opening angles. However, for 3D shallow cracks ($b/a=7$), the 3D leading-term solution is relatively accurate, which indicates that the tip fields of shallow cracks are less sensitive to the higher-order term.

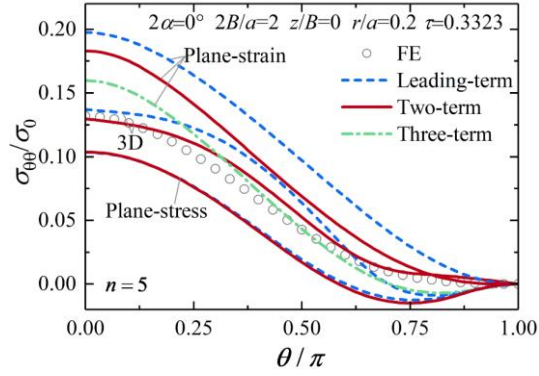

Fig. 18 Comparison of circumferential distribution of opening stress at the mid-plane of the 3D crack specimen

A comparison of circumferential distribution of opening stress is made as shown in Fig. 18. Based on the FE results, it is found that the stress state is between the plane-strain and plane-stress condition. According to the comparison, the leading-term and the two-term solutions under plane-strain and plane-stress condition are not accurate enough. And the plane-strain three-term and 3D leading-term solutions have better agreement with the FE results. The high-order terms are well considered in plane-strain three-term solution. And the 3D characteristics are well considered in the 3D leading-term solution in which $T_z$ is introduced as an internal feature. Herein, the 3D two-term solution combines their advantages and has good accuracy.

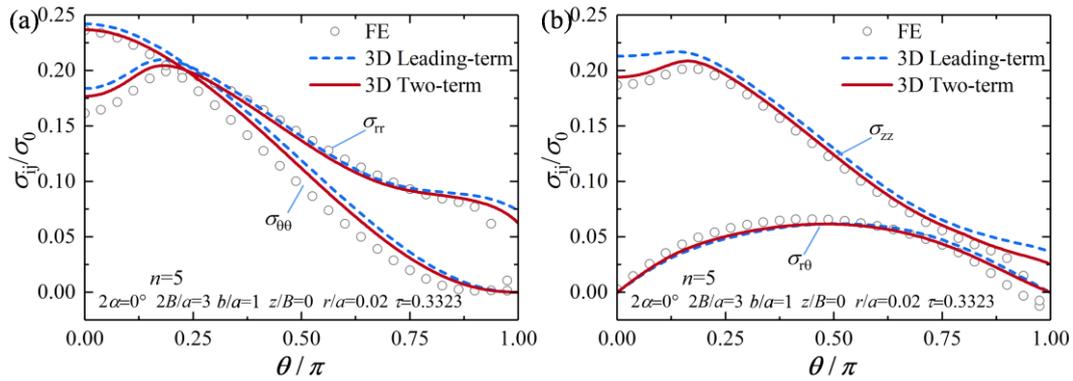

Fig. 19 Circumferential distribution of stress components in the 3D crack specimen when $r/a = 0.02$

It should be emphasized that the leading-term solution is the dominant term in the expansion of stress. As shown in Fig. 19, the 3D leading-term solution is accurate enough when the radial distance from the crack tip is small. This is the characteristic of asymptotic solution that the smaller the radial distance is, the less influences the higher-order term will make.

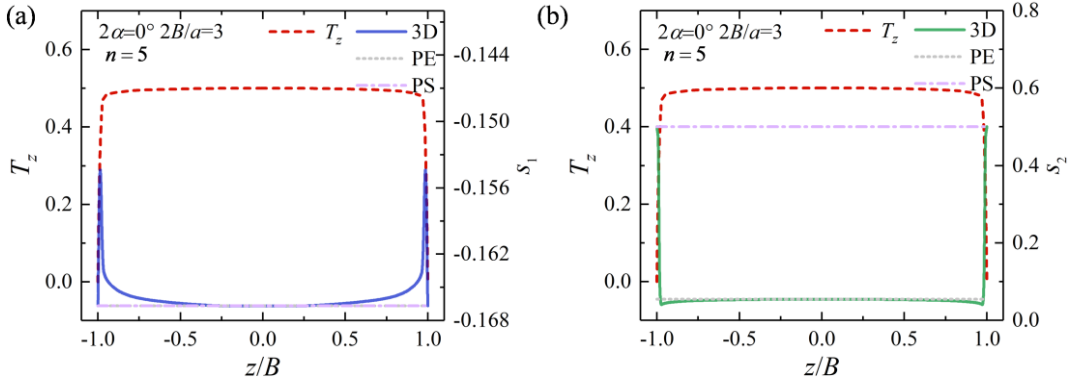

Fig. 20 The distributions of stress exponents and $T_z$ along the thickness at the crack front (a) first-order (b) second-order

Similar as the 3D notches in thick specimen, the distributions of stress exponents along the thickness at the crack front is also given in Fig. 20. The stress exponents change rapidly near the free surface.

## 5. On 3D constraint effect

### 5.1 Characteristics of 3D constraint effect

As presented in Section 3, the proposed $K_N^T - A_2^T$ and $C^T(t, T_z) - A_2^T$ solutions for 3D sharp V-notches and cracks show good agreement with the FEA results. It indicates that the higher-order solution based on the proposed 3D constraint effect for 3D sharp V-notches and cracks has significant advantages. To be specific, the $K_N^T - A_2^T$ solution has better accuracy at the area with a considerable radial distance away from the notch or crack tip. In addition, the $K_N^T - A_2^T$ solution is more accurate in the thin specimen and deep notches or cracks. In the author's opinion, the $K_N^T - A_2^T$ solution based on the 3D constraint effect is specially proposed for the 3D notches and cracks. It means that the out-of-plane effect on the leading term and the higher-order term are taken into consideration. Hence, the 3D constraint effect makes the out-of-plane effect and in-plane constraint effect for the 3D fracture problem form an undivided and unified theory.

In other words, the proposed $K_N^T - A_2^T$ solution provides a new perspective of constraints in 3D fracture problems. In previous literature, the constraint of 3D cracks and notches are divided into out-of-plane constraint and in-plane constraint (Matvienko et al., 2013; Qian et al., 2014; Shlyannikov et al., 2011; Sun et al., 2022). As shown in Fig. 21a, on the one hand, the out-of-plane factor $T_z$ affects the leading term of the tip fields of cracks or notches. Hence, $T_z$ is used to describe out-of-plane constraints directly (Cui and Guo, 2022; Xiang et al., 2011; Yi and Wang, 2020). On the other hand, $A_2$ is a parameter representing the constraint proposed for the 2D plane problems. Therefore, $A_2$ is taken as an appropriate parameter to represent in-plane constraint in 3D fracture problems spontaneously (Cui and Guo, 2020; Wang et al., 2014). Since $A_2$ is calculated based on the first-order leading term which is affected by $T_z$, $A_2$ contains the effect of $T_z$. However, the second-order stress exponents and angular distribution functions do not seem to be affected by $T_z$. In fact, it is reasonable when the out-of-plane effect or the constraint effect is not remarkable. For instance, the $A_2$-based theory is practical when the stress state approaches to the plane-strain or plane-stress cases for 3D notches or cracks. Moreover, related scholars put forward the concept of unified constraints in the experiment (Tonge et al., 2020) and application (Ma et al., 2016).

For 3D notches and cracks, the out-of-plane effect is of great importance for the constraint effect as presented in Section 3. The out-of-plane factor $T_z$ plays a significant role in both the leading term (Kong et al., 2022a) and the higher-order term. The two parts determine the level and effect of the entire constraint (see Fig. 21b). Hence, the 3D constraint effect is very necessarily needed as it is hard to distinguish the in-plane and out-of-plane part. It fully considers the influence of the out-of-plane effect on higher-order terms. Although this paper demonstrates the advantage of the two-term solution under some conditions, the leading-term is still the most important term in the stress expansion. For large notch angle specimens, the advantage of higher order term solutions is not remarkable. However, combined leading and second terms will generally enlarge the fracture parameter dominant region and better quantify fracture resistance for shallow to deep cracks (small notch angle specimens) in engineering structures.

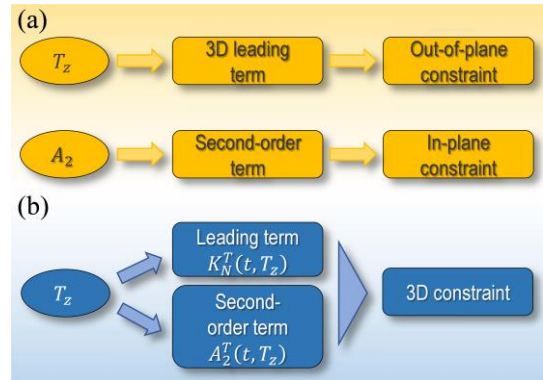

Fig. 21 About the constraint effect on 3D tip fields

To be specific, the characteristics of 3D constraint can be illustrated in terms of $A_2$ calculated under different assumptions. It is widely accepted that $A_2$ is contour-independent near the notch tip or crack tip under extensive creep(Chao et al., 2001; Dai et al., 2021). Consequently, the $A_2$ calculated from the FEA should approximately keep constant near the notch tip. Fig. 22 presents the $A_2$ calculated under different assumptions. The 3D notches with $2\alpha = 30°$ and $2\alpha = 120°$ are analyzed. If 2D plane $A_2$-based theory is directly applied without considering the out-of-plane effect, corresponding $A_2$ can be obtained for 3D finite element analysis results. The $A_2$ calculated based on the plane-stress hypothesis can be considered invalidly for thick specimens. The $A_2$ calculated based on the plane-strain hypothesis is also invalid for notches with small opening angles whose constraint effect may be significant. However, as shown in Fig. 22b, the $A_2$ calculated based on the plane-strain hypothesis is approximatively effective for notches with large opening angles whose constraint effect is not remarkable. On the contrary, the $A_2^{\mathrm{T}}$ calculated based on the 3D constraint effect in which the out-of-plane effect is considered in higher-order terms can always keep constant near the notch tip for different specimens.

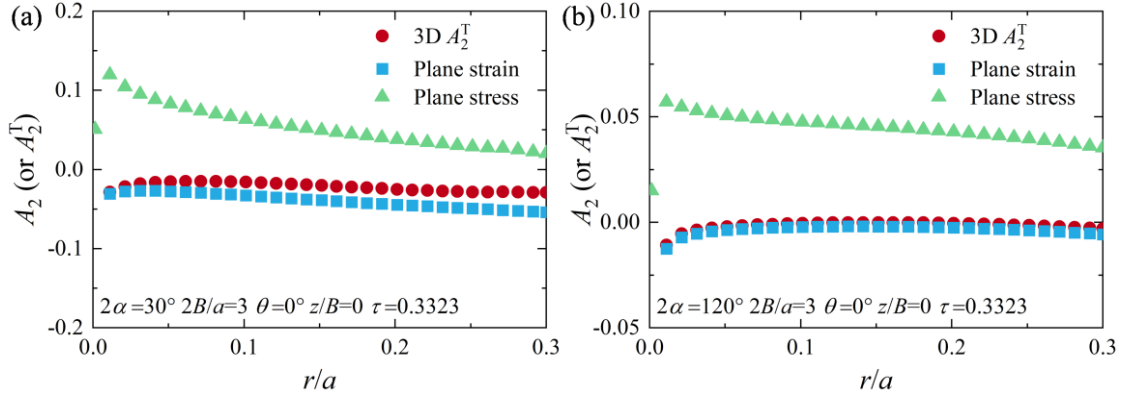

Fig. 22 The radial distribution of $A_2$ (or $A_2^T$) calculated with different stress state hypotheses for 3D sharp V-notches

Even though $A_2^T$ keeps constant along the radial direction when $\theta = 0°$, the second-order term is still related to $T_z$. Except for $A_2^T$, the stress exponent and angular distribution function are also the characteristics of the second-order term. As shown in Fig. 3 and Fig. 5~Fig. 6, the stress exponent and angular distribution function changes with $T_z$. Moreover, for different specimen with different thicknesses, the stress exponent also differs from each other (see Fig. 13 and Fig. 14). Different from the plane-strain and plane-stress cases whose characteristic parameter is invariable, the variation of 3D $s_2$ make $A_2^T$ tend to be stable on the specific slice parallel to the mid-plane of a specific specimen. Moreover, $A_2^T$ is an appropriate parameter to characterize constraint effect as it keeps constant on the mid-plane of a specific specimen.

Furthermore, the results obtained based on the 3D constraint effect show a different characteristic from the 2D plane cases. As discussed in Section 2, the condition in which the second-order term is not affected by the elastic part of the constitutive equation is defined as Case I. And the condition in which the second-order term is affected by the elastic part of the constitutive equation is defined as Case II. It is found that the distribution of Case I and Case II is significantly affected by the out-of-plane factor $T_z$. Moreover, Case I and Case II can co-exist in one 3D notch or crack specimen. This phenomenon is distinguished from the 2D plane cases. Case I and Case II cannot co-exist in 2D plane notches or cracks. Case II only appears for specific ranges of material properties (Xia et al., 1993) and notch opening angles (Dai et al., 2021).

## 5.2 3D fracture parameters

As $A_2^{\text{T}}$ is a constant on the mid-plane of a specific specimen, it is an appropriate parameter to characterize constraint effect which is dependent on the geometry of different specimens. According to the 3D constraint effect, the second-order term whose amplitude is characterized by $A_2^{\text{T}}$ is sensitive to the $T_z$. The influence of the out-of-plane factor $T_z$ on $A_2^{\text{T}}$ for 3D sharp V-notches with different ligament widths is shown in Fig. 23. The sphere marks present the results obtained from the FEA. The surface in Fig. 23 is the fitting surface. It is found that $A_2^{\text{T}}$ is related to the width of the ligament and the out-of-plane factor for a given notch opening angle ($2\alpha = 30°$ in Fig. 23). Similar to 2D plane cases, $A_2^{\text{T}}$ is an appropriate parameter to characterize the constraint level (Chao et al., 1994). The larger the $-A_2^{\text{T}}$ is, the higher the constraint level will be. Hence, the 3D constraint level increases with the increasing of out-of-plane factor $T_z$ and decreasing of $b/a$. In other words, the 3D constraint effect of deep notches is more pronounced than that of shallow notches.

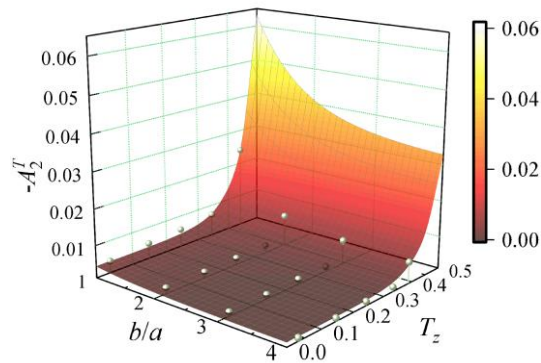

Fig. 23 The out-of-plane effect on $A_2^{\text{T}}$ for 3D sharp V-notches with different ligament width when $2\alpha = 30°$ and $n = 5$

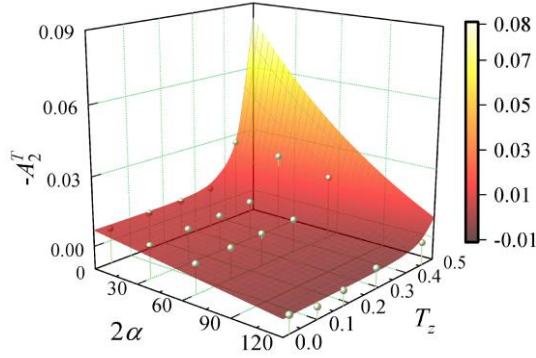

Fig. 24 The out-of-plane effect on $A_2^{\mathrm{T}}$ for 3D sharp V-notches with different opening angles when $b/a=1$ and $n=5$

Furthermore, the second-order term is also influenced by the notch opening angle. As shown in Fig. 24, $A_2^{\mathrm{T}}$ increases with the improvement of $T_z$ and decrease of the notch opening angle $2\alpha$. It is noted that the constraint effect is more remarkable for cracked and small notch angle specimens. According to Fig. 23 and Fig. 24, the larger the out-of-plane factor is, the more pronounced the 3D constraint effect will be. It indicates that the 3D constraint effect is most significant for the plane-strain condition. For 3D specimens with specific opening angle and $b/a$, the thicker the specimen is, the higher the 3D constraint effect will be. This can be concluded from the comparison of Fig. 8a, b and Fig. 9 that the difference between the two-term solution and the leading-term solution is much more pronounced for the thick specimen. In general, the 3D constraint level is apparently influenced by $T_z$. Hence, $T_z$ and $A_2^{\mathrm{T}}$ are combined to characterize 3D constraint effect. The results in Fig. 24 also imply that for large angle notches and thin specimens, the role of the 3D second-order term is limited. Under this condition, the 3D leading-term solution can have enough accuracy in describing the stress field.

## 6. Conclusions

This paper proposes a novel solution of mode I tip fields for 3D sharp V-notch in power-law creeping solids closing to the truth theoretically by introducing $T_z$ into the higher-order governing equations, which can be also degenerated to crack cases. The 3D constraint effect for 3D fracture issues is proposed based on the solution. Moreover, the characteristics of 3D constraint are also discussed. According to the theoretical analysis and numerical verification, the conclusions are drawn as follows:

1) The higher-order terms of the tip fields considering the out-of-plane effect for 3D sharp V-notches and cracks in power-law creeping materials are present. The stress exponents and angular distribution functions of the second-order term for 3D sharp V-notches and cracks are given. The second-order stress exponents and angular distribution functions are found to be sensitive to the out-of-plane factor $T_z$. Moreover, different from 2D plane conditions, the second-order term can be related to the elastic part of the constitutive equations with decreasing $T_z$ and increasing notch angle $2\alpha$ for specific material properties.

2) A higher-order termed tip field solution of 3D sharp V-notch with 3D constraint effect, i.e., $K_N^T - A_2^T$ solution, is proposed and verified by FE analyses. The solution can naturally be degenerated to $C^T(t, T_z) - A_2^T$ solution for 3D cracks. Compared with the 2D leading-term and two-term and three-term solutions and the 3D leading-term solution, the $K_N^T - A_2^T$ solution shows better agreement with the FE results for 3D sharp V-notches (especially for notches with small angles and ligament width) and cracks. The $K_N^T - A_2^T$ solution gives the accurate angular distribution of all the in-plane stress components and out-of-plane normal stress $\sigma_{zz}$. Similarly, unlike 2D plane solutions, the $K_N^T - A_2^T$ solution can contain both Case I and II in one specimen.

3) The 3D constraint effect for notches and cracks is analyzed through combining $T_z$ and $A_2^T$. Based on the 3D higher-order solution, it is verified that the in-plane and out-of-plane parts of constraint effect is highly interlinked. A theoretical framework for analyzing the 3D fracture issues with constraint effect is provided. The 3D constraint effect analysis based on the 3D two-order term solution is rigorously asymptotic solution. Out-of-plane factor $T_z$ influences both the leading term and the higher-order terms. $T_z$ and $A_2^{\text{T}}$ are combined to characterize 3D constraint effect. It should be noted that $A_2^{\text{T}}$ is influenced by $T_z$ as the leading term and the second-order term are both related to $T_z$. The 3D constraint effect becomes more apparent for increasing $T_z$, decreasing notch angle $2\alpha$ and the relative width of ligament $b/a$.

Although the combination of leading-term and second-order term shows better prediction accuracy for tip fields of 3D specimens with small notch angles (including crack) in regions far from the tip, it is noted that the 3D leading term solution is still the dominant term than other terms due to the asymptotic nature. For specimens with large notch angle, the 3D leading term is accurate enough to characterize the 3D tip fields. The tip field solution and 3D constraint effect proposed in this paper may enhance the understanding of constraint effects in 3D fracture issues. It may provide a more accurate stress analysis framework and reasonable characteristic parameters of constraint effect for the evaluation of 3D creep notches and cracks.

## Acknowledgment

The study was supported by the Major Project of the National Natural Science Foundation of China (12090033), and partly by National Natural Science Foundation of China (12272012).

## Appendix A: Analysis of stress exponents

The stress exponent characteristics of 3D sharp V-notches and cracks considering the out-of-plane effect has been analyzed in our previous work (see Appendix A in Kong et al. (2022a)). For the convenience of reading, the details are shown as follows:

The characteristics of $T_z$ are totally same as those proposed by Guo (1993a) when analyze the singular structure of the elastoplastic tip field. And the basic hypotheses mentioned by Guo (1993a) are also kept in the following analysis on the higher-order term. The details are discussed comprehensively in Section 2.3 and Section 3.2 in the original paper (Guo, 1993a).

At the beginning, the stress function is employed for the analysis. The stress function is written as:

$$\chi_1 = \sum_{\beta} D_{\beta}(t) r^{\lambda_1^{(\beta)}(T(z,t))} \tilde{\chi}_1^{(\beta)}(\theta, T_z)$$

$$\chi_2 = \sum_{\beta} D_{\beta}(t) r^{\lambda_2^{(\beta)}(T(z,t))} \tilde{\chi}_2^{(\beta)}(\theta, T_z) \quad \text{(A.1)}$$

$$\chi_3 = \sum_{\beta} D_{\beta}(t) r^{\lambda_3^{(\beta)}(T(z,t))} \tilde{\chi}_3^{(\beta)}(\theta, T_z)$$

in which $D_{\beta}$, $\lambda_i^{(\beta)}$, $\tilde{\chi}_i^{(\beta)}$ denote amplitude, stress function exponent and angular distribution function, respectively.

Based on the relation between Cartesian and cylindrical coordinates:

$$x = r\cos\theta, y = r\sin\theta, z = z \quad \text{(A.2)}$$

It is obtained that:

$$\frac{\partial r}{\partial x} = \cos\theta, \frac{\partial r}{\partial y} = \sin\theta, \frac{\partial \theta}{\partial x} = -\frac{\sin\theta}{r}, \frac{\partial \theta}{\partial y} = \frac{\cos\theta}{r} \quad \text{(A.3)}$$

Combining Eq. (5) and Eq. (A.1), the stress components are deduced:

$$r^{s_\beta^{xy}} \tilde{\sigma}_{xy}^{(\beta)} = -\frac{\partial^2 \chi_3}{\partial x \partial y}$$

$$= -\left\langle r^{\lambda_3^{(\beta)}-2} \left\{ \left[ \lambda_3^{(\beta)}(\lambda_3^{(\beta)}-2)\tilde{\chi}_3^{(\beta)} - \frac{\partial^2 \tilde{\chi}_3^{(\beta)}}{\partial \theta^2} - \frac{\partial^2 \tilde{\chi}_3^{(\beta)}}{\partial T_z \partial \theta} \frac{\partial T_z}{\partial \theta} - \frac{\partial \tilde{\chi}_3^{(\beta)}}{\partial T_z} \frac{\partial^2 T_z}{\partial \theta^2} \right] \cos\theta \sin\theta \right.\right.$$

$$-(\lambda_3^{(\beta)}-1)\left( \frac{\partial \tilde{\chi}_3^{(\beta)}}{\partial \theta} + \frac{\partial \tilde{\chi}_3^{(\beta)}}{\partial T_z} \frac{\partial T_z}{\partial \theta} \right) \sin^2\theta$$

$$+\left( \lambda_3^{(\beta)} \frac{\partial \tilde{\chi}_3^{(\beta)}}{\partial \theta} - \frac{\partial \tilde{\chi}_3^{(\beta)}}{\partial \theta} - \frac{\partial \tilde{\chi}_3^{(\beta)}}{\partial T_z} \frac{\partial T_z}{\partial \theta} \right) \cos^2\theta \bigg\}$$

$$+ r^{\lambda_3^{(\beta)}-1} \left\{ (\lambda_3^{(\beta)}-1) \frac{\partial \tilde{\chi}_3^{(\beta)}}{\partial T_z} \frac{\partial T_z}{\partial r} \cos(\theta)\sin(\theta) \right.$$

$$\left. -\left[ \frac{\partial}{\partial T_z}\left( \frac{\partial \tilde{\chi}_3^{(\beta)}}{\partial \theta} \right) \frac{\partial T_z}{\partial r} + \frac{\partial \tilde{\chi}_3^{(\beta)}}{\partial T_z} \frac{\partial^2 T_z}{\partial r \partial \theta} \right] \left[ \sin^2\theta - \cos^2\theta \right] \right\}$$

$$\left. + r^{\lambda_3^{(\beta)}} \frac{\partial \tilde{\chi}_3^{(\beta)}}{\partial T_z} \frac{\partial^2 T_z}{\partial r^2} \cos\theta \sin\theta \right\rangle$$

(A.4)

$$r^{s_\beta^{yz}} \tilde{\sigma}_{yz}^{(\beta)} = -\frac{\partial^2 \chi_1^{(\beta)}}{\partial y \partial z}$$

$$= -\left\langle r^{\lambda_1^{(\beta)}-1} \left[ \left( \frac{\partial \lambda_1^{(\beta)}}{\partial z} \tilde{\chi}_1^{(\beta)} + \lambda \frac{\partial \tilde{\chi}_1^{(\beta)}}{\partial T_z} \frac{\partial T_z}{\partial z} \right) \sin(\theta) + \frac{\partial \tilde{\chi}_1^{(\beta)}}{\partial T_z} \frac{\partial^2 T_z}{\partial \theta \partial z} \cos(\theta) \right] \right.$$

$$+ r^{\lambda_1^{(\beta)}-1} \ln(r) \left[ \lambda \frac{\partial \lambda_1^{(\beta)}}{\partial z} \tilde{\chi}_1^{(\beta)} \sin(\theta) + \frac{\partial \lambda_1^{(\beta)}}{\partial z} \left( \frac{\partial \tilde{\chi}_1^{(\beta)}}{\partial \theta} + \frac{\partial \tilde{\chi}_1^{(\beta)}}{\partial T_z} \frac{\partial T_z}{\partial \theta} \right) \cos(\theta) \right]$$

$$\left. + r^{\lambda_1^{(\beta)}} \frac{\partial \tilde{\chi}_1^{(\beta)}}{\partial T_z} \frac{\partial^2 T_z}{\partial r \partial z} \sin(\theta) + r^{\lambda_1^{(\beta)}} \ln(r) \frac{\partial \lambda_1^{(\beta)}}{\partial z} \frac{\partial \tilde{\chi}_1^{(\beta)}}{\partial T_z} \frac{\partial T_z}{\partial r} \sin(\theta) \right\rangle$$

(A.5)

in which $s_{xy}^{(\beta)}$ and $s_{yz}^{(\beta)}$ are the stress exponents in the stress hierarchy.

It should be noted that the inequality about the exponent of stress function has been proven by Guo (1993a), i.e.,

$$\lambda_1^{(1)}(z) > 1 \tag{A.6}$$

And as higher-order term discussed in the present work, it is natually concluded that

$$\lambda_1^{(\beta)} \geq \lambda_1^{(1)} > 1 \tag{A.7}$$

The inequality mentioned above can guarantee the order of $r^{\lambda_1^{(\beta)}-1}\ln(r)$ and $r^{\lambda_1^{(\beta)}}\ln(r)$ in Eq. (A.5) can be obtained according to L'Hôpital's rule.

According to Eq. (10), $T_z$ can be expressed as:

$$T_z(r,\theta,z,t) = T(z,t) + \sum_{1}^{n} b_i(\theta,z,t) r^{\eta_i} \tag{A.8}$$

As $T_z$ is a finite value, it contains the constant term $T(z,t)$, and $\eta_i \geq 0$. Herein, after ignoring the higher-order terms, it is concluded that:

$$\begin{aligned} s_\beta^{xy} &= \lambda_3^{(\beta)} - 2 \\ s_\beta^{yz} &= \lambda_1^{(\beta)} - 1 \end{aligned} \tag{A.9}$$

Similarly,

$$s_\beta^{xz} = \lambda_2^{(\beta)} - 1 \tag{A.10}$$

For the normal stress components, the stress exponents are deduced by substituting Eq. (A.1) into Eq. (5) that

$$\begin{aligned} r^{s_\beta^{xx}} \tilde{\sigma}_{xx}^{(\beta)} &= a_{xx}^{(\beta)} \cdot r^{\lambda_3^{(\beta)}(T(z,t))-2} + b_{xx}^{(\beta)} \cdot r^{\lambda_2^{(\beta)}(T(z,t))} \\ r^{s_\beta^{yy}} \tilde{\sigma}_{yy}^{(\beta)} &= a_{yy}^{(\beta)} \cdot r^{\lambda_3^{(\beta)}(T(z,t))-2} + b_{yy}^{(\beta)} \cdot r^{\lambda_1^{(\beta)}(T(z,t))} \\ r^{s_\beta^{zz}} \tilde{\sigma}_{zz}^{(\beta)} &= a_{zz}^{(\beta)} \cdot r^{\lambda_2^{(\beta)}(T(z,t))-2} + b_{zz}^{(\beta)} \cdot r^{\lambda_1^{(\beta)}(T(z,t))-2} \end{aligned} \tag{A.11}$$

According to Eq. (9) and the finiteness of $T_z$, the out-of-plane normal stress component is of the same order as the in-plane normal stress components. Hence, it is assumed that

$$\lambda_1^{(\beta)} - 2 = \lambda_2^{(\beta)} - 2 = \lambda_3^{(\beta)} - 2 \tag{A.12}$$

Thus, it is obtained that:

$$\lambda_1^{(\beta)} = \lambda_2^{(\beta)} = \lambda_3^{(\beta)} = \lambda^\beta(T(z,t)) \tag{A.13}$$

Accordingly,

$$\begin{aligned} s_\beta^{xx} &= s_\beta^{yy} = s_\beta^{xy} = s_\beta^{zz} = \lambda^{(\beta)} - 2 = s_\beta \\ s_\beta^{xz} &= s_\beta^{yz} = \lambda^{(\beta)} - 1 = s_\beta + 1 \end{aligned} \tag{A.14}$$

where $s_\beta$ is the exponent of in-plane stress components and out-of-plane stress component $\sigma_{zz}$. Hence, it is concluded that:

$$\begin{cases} s_1^{xx} = s_1^{yy} = s_1^{xy} = s_2^{zz} = s_1 \\ s_1^{xz} = s_1^{yz} = s_1 + 1 \end{cases} \quad (A.15)$$

and

$$\begin{cases} s_2^{xx} = s_2^{yy} = s_2^{xy} = s_2^{zz} = s_2 \\ s_2^{xz} = s_2^{yz} = s_2 + 1 \end{cases} \quad (A.16)$$

Hence, the second-order part of the out-of-plane shear stress is the higher-order term which can be neglected in the analysis. However, it is necessary to prove $\left(s_1^{xz} = s_1^{yz}\right) > \left(s_2^{xx} = s_2^{yy} = s_2^{xy} = s_2^{zz}\right)$ to make sure the first-order part of the out-of-plane shear stress will not influence the establishment of the second-order governing equation. In other words, only if $s_1 + 1 > s_2$, the first-order part of the out-of-plane shear stress can be ignored in the second-order analysis. The proof process is as follows.

For second-order part of the in-plane stress components, it has been widely accepted that

$$s_2 \leq (2-n)s_1 \quad (A.17)$$

for both notches (Dai et al., 2021) and cracks (Chao et al., 2001). For 3D notches and cracks, it is concluded that

$$s_1 \geq -\frac{1}{n+1} \quad (A.18)$$

as the minimum first-order stress exponent of cracks and cracks is $-1/(n+1)$ according to the previous literature (Kong et al., 2022a; Xiang et al., 2011).

For most of the creep solids, $n > 1$, hence,

$$(1-n)s_1 \leq \frac{n-1}{n+1} < 1 \quad (A.19)$$

Therefore, it is obtained that

$$(2-n)s_1 < s_1 + 1 \quad (A.20)$$

Combining Eq. (A.17) and Eq. (A.20), the conclusion is drawn as below:

$$s_2 < s_1 + 1 \quad (A.21)$$

In general, $\sigma_{xz}$ and $\sigma_{yz}$ are the higher-order terms which can be ignored in the first and second-order asymptotic analysis. Therefore, in the second-order asymptotic analysis, all the desired stress components can be determined by $\chi_3$.

## Appendix B: Finite element model

The FEM is employed to verify the accuracy of the proposed tip field $K_N^T - A_2^T$ solution. The numerical simulation is performed by using FEM software ABAQUS. The power-law creeping behavior defined in Eq. (1) is considered. The specific material superalloy Inconel 800H is selected for the simulation, and the material properties at 649°C (Chao et al., 2001) are shown in Table A. 1. The 3D single-edged sharp V-notches with different opening angles are considered in the calculation. The 3D FEA meshes are shown in Fig A. 1. The meshes are refined near the notch tip. The used mesh size near the notch tip is about 5 μm. The mesh convergence is verified. The notch depth $a$ is set as 5 mm. And specimens with different opening angles ($2\alpha = 30°$, $2\alpha = 120°$) and thicknesses ($2B/a = 3$, $2B/a = 0.1$) are calculated and analyzed in this section. The element type is set as C3D8R. And the deep notches with $a/b = 1$ are adopted to present the constraint effect.

Table A. 1 Material properties used in the calculation (Chao et al., 2001)

| Creep exponent | Young's modulus | Reference stress | Poisson's ratio | Creep coefficient | Reference strain rate |
|---|---|---|---|---|---|
| $n=5$ | 153606MPa | 416.8MPa | 0.33 | $1.352 \times 10^{-16}$ | $1.70065 \times 10^{-3}$ |

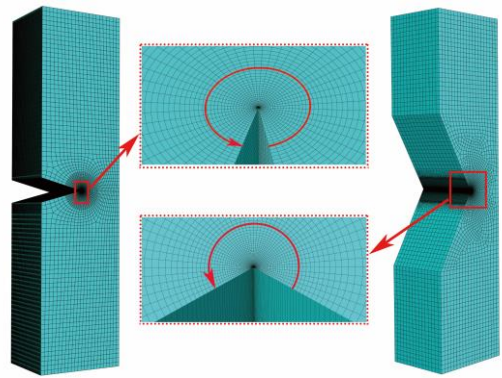

Fig A. 1 FEM meshes used in the simulation

The method proposed by Dai et al. (2021) to obtain the $K_N^T$ and $A_2^T$ from FEA results is used reasonably and practically in the present calculation.

# References

Ainsworth, R.A., 2006. R5 procedures for assessing structural integrity of components under creep and creep-fatigue conditions. Int Mater Rev 51, 107-126.

Budden, P.J., Ainsworth, R.A., 1999. The effect of constraint on creep fracture assessments. Int J Fracture 97, 237-247.

Chao, Y.J., Lam, P.-S., 2009. Constraint Effect in Fracture—What is It?, Proceeding of the 12th Intermational Conference of Fracture. Citeseer, Ottawa, Canada.

Chao, Y.J., Yang, S., Sutton, M.A., 1994. On the Fracture of Solids Characterized by One or 2 Parameters - Theory and Practice. Journal of the Mechanics and Physics of Solids 42, 629-647.

Chao, Y.J., Zhu, X.K., Zhang, L., 2001. Higher-order asymptotic crack-tip fields in a power-law creeping material. Int J Solids Struct 38, 3853-3875.

Cui, P., Guo, W., 2019. Higher order J-Tz-AT solution for three-dimensional crack border fields in power-law hardening solids. Eng Fract Mech 222, 106736.

Cui, P., Guo, W., 2022. A predicting model for three-dimensional crack growth in power-law creeping solids. Journal of the Mechanics and Physics of Solids 168, 105029.

Cui, P.F., Guo, W.L., 2020. Higher order C(t)-T-z-A(T) solution for three-dimensional creep crack border fields. Eng Fract Mech 236.

Dai, Y., Liu, Y., Qin, F., Chao, Y.J., Chen, H., 2020a. Constraint modified time dependent failure assessment diagram (TDFAD) based on C(t)-A2(t) theory for creep crack. Int J Mech Sci 165, 105193.

Dai, Y.W., Liu, Y.H., Qin, F., Chao, Y.J., Berto, F., 2019. Estimation of stress field for sharp V-notch in power-law creeping solids: An asymptotic viewpoint. Int J Solids Struct 180, 189-204.

Dai, Y.W., Liu, Y.H., Qin, F., Chao, Y.J., Qian, G.A., 2020b. C(t) dominance of the mixed I/II creep crack: Part II. Extensive creep. Theor Appl Fract Mec 106.

Dai, Y.W., Qin, F., Liu, Y.H., Chao, Y.J., 2021. On the second order term asymptotic solution for sharp V-notch tip field in elasto-viscoplastic solids. Int J Solids Struct 217, 106-122.

Davies, C.M., Dean, D.W., Yatomi, M., Nikbin, K.M., 2009. The influence of test duration and geometry on the creep crack initiation and growth behaviour of 316H steel. Mat Sci Eng a-Struct 510-11, 202-206.

Gallo, P., Berto, F., Glinka, G., 2016. Analysis of creep stresses and strains around sharp and blunt V-notches. Theor Appl Fract Mec 85, 435-446.

Guo, W., 1993a. Elastoplastic three dimensional crack border field—I. Singular structure of the field. Eng Fract Mech 46, 93-104.

Guo, W., 1993b. Elastoplastic three dimensional crack border field—II. Asymptotic solution for the field. Eng Fract Mech 46, 105-113.

Guo, W., 1995. Elasto-plastic three-dimensional crack border field—III. Fracture parameters. Eng Fract Mech 51, 51-71.

Hutchinson, J.W., 1968. Singular behaviour at the end of a tensile crack in a hardening material. Journal of the Mechanics and Physics of Solids 16, 13-31.

James, P., Coon, D., Austin, C., Underwood, N., Chevalier, M., Dean, D., 2020. Overview of Easics Validation Experiments and Code Comparison of R5, Rcc-Mrx and Asme Iii Division V. Proceedings of the Asme 2020 Pressure Vessels & Piping Conference (Pvp2020), Vol 1.

Kong, W.C., Dai, Y.W., Liu, Y.H., 2022a. Out-of-plane effect on the sharp V-notch tip fields in power-law creeping solids: A three-dimensional asymptotic analysis. Int J Solids Struct 236.

Kong, W.C., Dai, Y.W., Liu, Y.H., 2022b. Three-dimensional sharp V-notch stress intensity factor and strain energy rate density under creeping conditions. Eng Fract Mech 272.

Kuang, Z.B., Xu, X.P., 1987. Stress and Strain Fields at the Tip of a Sharp V-Notch in a Power-Hardening Material. Int J Fracture 35, 39-53.

Kwon, S.W., Sun, C.T., 2000. Characteristics of three-dimensional stress fields in plates with a through-the-thickness crack. Int J Fracture 104, 291-315.

Lazzarin, P., Zappalorto, M., 2012. A three-dimensional stress field solution for pointed and sharply radiused V-notches in plates of finite thickness. Fatigue Fract Eng M 35, 1105-1119.

Lazzarin, P., Zappalorto, M., Berto, F., 2015. Three-dimensional stress fields due to notches in plates under linear elastic and elastic-plastic conditions. Fatigue Fract Eng M 38, 140-153.

Lee, J., Keer, L., 1986. Study of a three-dimensional crack terminating at an interface. Journal of Applied Mechanics, Transactions ASME 53, 311-316.

Li, Y., Gong, B., Corrado, M., Deng, C., Wang, D., 2017. Experimental investigation of out-of-plane constraint effect on fracture toughness of the SE(T) specimens. Int J Mech Sci 128-129, 644-651.

Li, Z.H., Guo, W.L., Kuang, Z.B., 2000. Three-dimensional elastic stress fields near notches in finite thickness plates. Int J Solids Struct 37, 7617-7631.

Loghin, A., Joseph, P.F., 2020. Mixed mode fracture in power law hardening materials for plane stress. Journal of the Mechanics and Physics of Solids 139.

Ma, H.S., Wang, G.Z., Liu, S., Tu, S.T., Xuan, F.Z., 2016. In-plane and out-of-plane unified constraint-dependent creep crack growth rate of 316H steel. Eng Fract Mech 155, 88-101.

Matvienko, Y.G., Shlyannikov, V.N., Boychenko, N.V., 2013. In-plane and out-of-plane constraint parameters along a three-dimensional crack-front stress field under creep loading. Fatigue Fract Eng M 36, 14-24.

Meng, Q.H., Gao, Y., Shi, X.H., Feng, X.Q., 2022. Three-dimensional crack bridging model of biological materials with twisted Bouligand structures. Journal of the Mechanics and Physics of Solids 159.

Mostafavi, M., McDonald, S.A., Mummery, P.M., Marrow, T.J., 2013. Observation and quantification of three-dimensional crack propagation in poly-granular graphite. Eng Fract Mech 110, 410-420.

Mostafavi, M., Smith, D.J., Pavier, M.J., 2011. A micromechanical fracture criterion accounting for in-plane and out-of-plane constraint. Comp Mater Sci 50, 2759-2770.

Nakamura, T., Parks, D.M., 1990. Three-dimensional crack front fields in a thin ductile plate. Journal of the Mechanics and Physics of Solids 38, 787-812.

Nguyen, B.N., Onck, P.R., van der Giessen, E., 2000. On higher-order crack-tip fields in creeping solids. J Appl Mech-T Asme 67, 372-382.

O'Connor, A.N., Davies, C.M., Garwood, S.J., 2022. The influence of constraint on fracture toughness: Comparing theoretical T0 shifts in master curve analyses with experimental data. Eng Fract Mech 275, 108857.

Qian, G.A., Gonzalez-Albuixech, V.F., Niffenegger, M., 2014. In-plane and out-of-plane constraint effects under pressurized thermal shocks. Int J Solids Struct 51, 1311-1321.

Rice, J.R., 1989. Weight function theory for three-dimensional elastic crack analysis. ASTM International.

Rice, J.R., Rosengren, G.F., 1968. Plane Strain Deformation near a Crack Tip in a Power-Law Hardening Material. Journal of the Mechanics and Physics of Solids 16, 1-12.

Riedel, H., Rice, J., 1980. Tensile cracks in creeping solids. Fracture mechanics 12, 112-130.

Sharma, S.M., Aravas, N., 1991. Determination of higher-order terms in asymptotic elastoplastic crack tip solutions. Journal of the Mechanics and Physics of Solids 39, 1043-1072.

Shih, C., Moran, B., Nakamura, T., 1986. Energy release rate along a three-dimensional crack front in a thermally stressed body. Int J Fracture 30, 79-102.

Shlyannikov, V.N., Boychenko, N.V., Tartygasheva, A.M., 2011. In-plane and out-of-plane crack-tip constraint effects under biaxial nonlinear deformation. Eng Fract Mech 78, 1771-1783.

Sun, X.Y., Liu, Z., Wang, X., Chen, X., 2022. Determination of ductile fracture properties of 16MND5 steels under varying constraint levels using machine learning methods. Int J Mech Sci 224.

Tan, J.P., Tu, S.T., Wang, G.Z., Xuan, F.Z., 2013. Effect and mechanism of out-of-plane constraint on creep crack growth behavior of a Cr-Mo-V steel. Eng Fract Mech 99, 324-334.

Tan, J.P., Wang, G.Z., Tu, S.T., Xuan, F.Z., 2014. Load-independent creep constraint parameter and its application. Eng Fract Mech 116, 41-57.

Tonge, S.M., Simpson, C.A., Reinhard, C., Connolley, T., Sherry, A.H., Marrow, T.J., Mostafavi, M., 2020. Unifying the Effects of in and out-of-plane constraint on the fracture of ductile materials. Journal of the Mechanics and Physics of Solids 141, 103956.

Wang, E.Y., Zhou, W.X., Shen, G.W., 2014. Three-dimensional finite element analysis of crack-tip fields of clamped single-edge tension specimens - Part II: Crack-tip constraints. Eng Fract Mech 116, 144-157.

Weaver, J., 1977. Three-dimensional crack analysis. Int J Solids Struct 13, 321-330.

Xia, L., Wang, T.C., Shih, C.F., 1993. Higher-Order Analysis of Crack Tip Fields in Elastic Power-Law Hardening Materials. Journal of the Mechanics and Physics of Solids 41, 665-687.

Xiang, M., Guo, W., 2013. Formulation of the stress fields in power law solids ahead of three-dimensional tensile cracks. Int J Solids Struct 50, 3067-3088.

Xiang, M.J., Yu, Z.B., Guo, W.L., 2011. Characterization of three-dimensional crack border fields in creeping solids. Int J Solids Struct 48, 2695-2705.

Yang, S., Chao, Y.J., Sutton, M.A., 1993a. Complete Theoretical-Analysis for Higher-Order Asymptotic Terms and the Hrr Zone at a Crack Tip for Mode-I and Mode-Ii Loading of a Hardening Material. Acta Mech 98, 79-98.

Yang, S., Chao, Y.J., Sutton, M.A., 1993b. Higher-Order Asymptotic Crack-Tip Fields in a Power-Law Hardening Material. Eng Fract Mech 45, 1-20.

Yi, D.K., Wang, T.C., 2020. The effect of out-of-plane constraint on the stress fields near the front of a crack in a thin ductile plate. Int J Solids Struct 190, 244-257.

Zhao, L., Wu, Y., Xu, L., Han, Y., 2023. Characterization of creep constraint effects on creep crack growth behavior by Q-type parameters. Eng Fract Mech 279, 109015.

Zhu, H.L., Xu, J.Q., Feng, M.L., 2011. Singular fields near a sharp V-notch for power law creep material. Int J Fracture 168, 159-166.

Zhu, X.K., Chao, Y.J., 1999. Characterization of constraint of fully plastic crack-tip fields in non-hardening materials by the three-term solution. Int J Solids Struct 36, 4497-4517.

Zhu, X.K., Liu, G.T., Chao, Y.J., 2001. Three-dimensional stress and displacement fields near an elliptical crack front. Int J Fracture 109, 383-401.